\documentclass[10pt]{article}

\usepackage{amsmath}

\usepackage{array}

\usepackage{appendix}

\usepackage{tocloft}                   

\usepackage{graphicx}

\usepackage{amsfonts}

\usepackage{amssymb}

\usepackage{mathrsfs}

\usepackage{yfonts}

\usepackage{euscript}

\usepackage{centernot}                 

\usepackage{ifsym}                     

\usepackage{upgreek}

\usepackage{mathtools}

\usepackage{color}

\usepackage{slantsc}
\usepackage{calligra}

\usepackage{bbold}          

\usepackage[T1]{fontenc}

\usepackage{epsf}

\usepackage{latexsym}

\usepackage{tipa}

\usepackage{makeidx}

\makeindex

\def\b{\begin{equation}}

\def\e{\begin{equation}}

\def\be{\begin{equation}}              

\def\ee{\end{equation}}

\def\beq{\begin{equation}}

\def\eeq{\end{equation}}

\def\bea{\begin{eqnarray}}

\def\eea{\end{eqnarray}}

\def\m{\mbox{ }}

\def\mma {\m , \m \m }

\def\!{\hspace{-1.6667em}}

\def\c{\cite}

\def\n{\noindent}

\def\u{\underline}

\def\w{\widetilde}

\def\s{\stackrel}

\def\slLambda{\mathit{\Lambda}}                   

\def\bigg{\mbox{\boldmath$g$}}

\def\sbig{\mbox{\scriptsize\boldmath$g$}}

\def\biq{\mbox{\boldmath$q$}}

\def\biP{\mbox{\boldmath$P$}}

\def\biQ{\mbox{\boldmath$Q$}}

\def\mD{\mbox{D}}                        

\def\mF{\mbox{F}}

\def\mI{\mbox{I}}                        

\def\mJ{\mbox{J}}  

\def\mK{\mbox{K}}

\def\mM{\mbox{M}}                        

\def\mN{\mbox{N}}

\def\mX{\mbox{X}}

\def\mY{\mbox{Y}}

\def\mZ{\mbox{Z}}

\def\me{\mbox{e}}

\def\mg{\mbox{g}}

\def\mh{\mbox{h}}

\def\mp{\mbox{p}}

\def\ms{\mbox{s}}

\def\mt{\mbox{t}}

\def\uq{\u{\mbox{q}}}

\def\ux{\u{\mbox{x}}}

\def\bh{\u{\u{\mbox{h}}}  }            

\def\bC{\mbox{\bf C}}                    

\def\bG{\mbox{\bf G}}                     

\def\bg{\mbox{\bf g}}

\def\bh{\mbox{\bf h}}

\def\bp{\mbox{\bf p}}

\def\buu{\mbox{\bf u}}             

\def\bupSigma{\mbox{\boldmath$\Sigma$}}                 
\def\sbupSigma{\mbox{\scriptsize\boldmath$\Sigma$}}     
\def\sbSigma{\mbox{\scriptsize\boldmath$\Sigma$}}

\def\cC{{\mathscr C}}

\def\sa{\mbox{\scriptsize a}}

\def\sb{\mbox{\scriptsize b}}

\def\se{\mbox{\scriptsize e}}

\def\si{\mbox{\scriptsize i}}

\def\sll{\mbox{\scriptsize l}}  

\def\sm{\mbox{\scriptsize m}}

\def\sn{\mbox{\scriptsize n}} 

\def\so{\mbox{\scriptsize o}}

\def\sr{\mbox{\scriptsize r}}

\def\sss{\mbox{\scriptsize s}}  

\def\st{\mbox{\scriptsize t}}

\def\su{\mbox{\scriptsize u}}

\def\sv{\mbox{\scriptsize v}}

\def\sw{\mbox{\scriptsize w}}

\def\sA{\mbox{\scriptsize A}} 

\def\sB{\mbox{\scriptsize B}}

\def\sD{\mbox{\scriptsize D}}

\def\sF{\mbox{\scriptsize F}}

\def\sG{\mbox{\scriptsize G}}

\def\sH{\mbox{\scriptsize H}}

\def\sM{\mbox{\scriptsize M}} 

\def\sN{\mbox{\scriptsize N}} 

\def\sO{\mbox{\scriptsize O}}

\def\sR{\mbox{\scriptsize R}}

\def\sS{\mbox{\scriptsize S}}

\def\sT{\mbox{\scriptsize T}}

\def\sW{\mbox{\scriptsize W}}

\def\sX{\mbox{\scriptsize X}}

\def\sfQ{\mbox{\sffamily{\scriptsize Q}}}      

\def\sbg{\mbox{{\bf \scriptsize g}}}

\def\sbcC{\mbox{\boldmath \scriptsize ${\cal C}$}}

\def\sbcF{\mbox{\boldmath \scriptsize ${\cal F}$}}

\def\sbcG{\mbox{\boldmath \scriptsize ${\cal G}$}}

\def\cr{\mbox{\scriptsize{\bf $\m  \times \m $}}}

\def\sumi2{\sum\mbox{}_{\mbox{}_{\mbox{\scriptsize $i$=1}}}^2}

\def\sumi3{\sum\mbox{}_{\mbox{}_{\mbox{\scriptsize $i$=1}}}^3}

\def\sumABcycles3{\sum\mbox{}_{\mbox{}_{\mbox{\scriptsize cycles $A,B$=1}}}^{3}}

\def\sumCDcycles3{\sum\mbox{}_{\mbox{}_{\mbox{\scriptsize cycles $C,D$=1}}}^{3}}

\def\sumj3{\sum\mbox{}_{\mbox{}_{\mbox{\scriptsize $j$=1}}}^3}

\def\sumk3{\sum\mbox{}_{\mbox{}_{\mbox{\scriptsize $k$=1}}}^3}

\def\prodiA1{\prod\mbox{}_{\mbox{}_{\mbox{\scriptsize $i$=1}}}^{A - 1}}

\def\bigtimes{\mbox{\Large $\times$}}

\def\d{\textrm{d}}                                                  

\def\pa{\partial}                                                   

\def\es{\m = \m}

\def\:={\m := \m}

\def\=:{\m =: \m}

\def\FrA{\mbox{$\mathfrak{A}$}}                                

\def\sFrA{\mbox{\scriptsize$\mathfrak{A}$}}                                

\def\FrT{\mathfrak{T}}                                         

\def\FrC{\mbox{$\mathfrak{C}$}}                                
                                                       
\def\FrS{\mbox{\Large $\mathfrak{s}$}}                         
\def\FrU{\mbox{$\mathfrak{U}$}}                                
\def\FrV{\mbox{$\mathfrak{V}$}}                                
\def\Frm{\mbox{\Large $\mathfrak{m}$}}                         
\def\FrM{\mbox{$\mathfrak{M}$}}                                
\def\FrN{\mbox{$\mathfrak{N}$}}                                
\def\lFrg{\mbox{\Large$\mathfrak{g}$}}                         
\def\nFrg{\mbox{\large$\mathfrak{g}$}}                         
\def\FrN{\mathfrak{N}}                                         
                                                  
\def\FrT{\mbox{\boldmath$\mathfrak{T}$}}                       
\def\Hilb{\mbox{{\boldmath$\mathfrak{H}$}ilb}}                 

\def\scC{\mbox{\scriptsize ${\cal C}$}}                    
\def\scE{\mbox{\scriptsize ${\cal E}$}}                    

\def\scF{\mbox{\scriptsize ${\cal F}$}}

\def\scH{\mbox{\scriptsize ${\cal H}$}}                    

\def\scM{\mbox{\scriptsize ${\cal M}$}}                    

\def\scS{\mbox{\scriptsize ${\cal S}$}}                    

\def\Flin{\scF\mbox{lin}}                                  

\def\bFlin{\sbcF\mbox{\bf lin}} 

\def\Chronos{\scC\mbox{hronos}}                            

\def\bGauge{\sbcG\mbox{\bf auge}}

\def\iS{\mbox{\scriptsize$S$}}                             

\def\sbiU{\mbox{\boldmath\scriptsize$U$}}                   %

\def\sbiC{\mbox{\boldmath\scriptsize$C$}}

\def\sbiK{\mbox{\boldmath\scriptsize$K$}} 

\def\sbiB{\mbox{\boldmath\scriptsize$B$}} 

\def\siB{\mbox{\scriptsize$B$}} 

\def\sbiD{\mbox{\boldmath\scriptsize$D$}} 

\def\FrQ{\mbox{\Large $\mathfrak{q}$}}                               
\def\bFrC{\mbox{\boldmath$\mathfrak{C}$}}                            
\def\Phase{\mbox{{\boldmath$\mathfrak{P}$}hase}}                     

\def\bFrR{\mbox{\boldmath$\mathfrak{R}$}}                            
\def\Rig-Phase{\bFrR\mbox{ig-}\Phase}                                
                                                                													   
\def\bFrM{\mbox{\boldmath${\mathfrak{M}}$}}                             

\def\bFrR{\mbox{\boldmath$\mathfrak{R}$}}                            

\def\bFrR{\mbox{\boldmath$\mathfrak{R}$}}                            

\def\1mat{\u{\u{1}}}                                                 

\def\Positive-Modespace{\mbox{{\boldmath$\mathfrak{M}$}odespace$^+$}}

\def\POSITIVE-MODESPACE{\mbox{{\boldmath$\mathfrak{M}$}ODESPACE$^+$}}
                                                                                                                             														
\def\bFrS{\mbox{\Large $\mathfrak{s}$}}                              
			
\def\blFrS{\mbox{\LARGE $\mathfrak{s}$}}                             
\def\Riem{\bFrR\mbox{iem}}                                           
\def\CRiem{\bFrC\Riem}                                               

\def\Superspace{\bFrS\mbox{uperspace}}                               
\def\lSuperspace{\blFrS\mbox{uperspace}}                             

\def\CS{\bFrC\bFrS}                                                  

\def\lattice{\mbox{\bf\Large$\mathfrak{L}$}}                                      

\def\sFrC{\mbox{\boldmath\scriptsize$\mathfrak{C}$}}                        

\def\sFrB{\mbox{\boldmath\scriptsize$\mathfrak{B}$}}                        

\def\Kin-Hilb{\mbox{{\boldmath$\mathfrak{K}$}in-\Hilb}}                     

\def\Mid-Hilb{\mbox{{\boldmath$\mathfrak{M}$}id-\Hilb}}                     

\def\Dyn-Hilb{\mbox{{\boldmath$\mathfrak{D}$}yn-\Hilb}}                     

\def\5Star{\mbox{\Large$\star$}}              

\def\Frr{\mbox{$\mathfrak{r}$}}

\begin{document}

\begin{titlepage}

\begin{center}

{\bf \large Shape Theory. III. Comparative Theory of Backgound Independence} 

\vspace{0.1in}

{\bf Edward Anderson}$^*$  

\end{center}

\begin{abstract}

Background Independence is the modern form of the relational side of the Absolute versus Relational (Motion) Debate. 
Difficulties with its implementation form the Problem of Time. 
Its 9 facets - Isham and Kucha\v{r}'s conceptual classification - correspond to 9 aspects of Background Independence 
as per the Author's classical-or-quantum theory-independent upgrade. 
8 are local. 
The most significant arena for these is brackets algebra, the first 5 involving canonical constraints as follows. 
1) Handling instantaneous gauge invariance. 
2) Resolving its apparent timelessness. 
3) Closure of 1-2)'s constraints. 
4) Expression in terms of observables: commutants with constraints. 
5) Reconstructing spacetime from constraint algebra rigidity. 
6) is the spacetime counterpart of 2-4), and 7) is spacetime's foliation independence.
8) handles nonuniqueness. 
9) renders 1-8) globally sound. 
We show how Shape(-and-Scale) Theory's mastery of 1) for $N$-point-particle models extends by placing a mechanics over shape(-and-scale) space to model 1-4). 
For flat-space Euclidean and similarity models, this gives a local resolution of the Problem of Time. 
This is moreover consistent within a global treatment if its reduced spaces are Hausdorff paracompact, which admit a Shrinking Lemma. 
GR's superspace is also Hausdorff paracompact. 
While 1-9) are poseable for all relativistic theories, and 1-4), 8), 9) for all theories 
-- to all levels of mathematical structure: affine, projective, conformal, topological manifold, topological space... -- resolution is on a case-by-case basis. 
Among $N$-point-particle theories, then, Article II's Hausdorff paracompact reduced space guarantee selects a very small subset. 
In particular, affine and projective shapes are precluded and conceiving in terms of shapes in space is preferred over doing so in spacetime. 
A substantial Selection Principle for Comparative Background Independence is thus born.
%

\end{abstract}

\n Physics keywords: Background Independence, configuration spaces, GR (conformal) superspace. 

\m

\n Maths keywords: Applied Geometry. Applied Topology. Lie's: derivative, bracket, algebra, flow, integral invariants. Rigidity.  
Stratified manifolds. Shape Theory. Hausdorff Paracompact spaces, Shrinking Lemma, (pre)sheaves. 

\m

\n PACS: 04.20.Cv, 02.40.-k  $^*$ dr.e.anderson.maths.physics *at* protonmail.com 

\vspace{-0.1in}

\section{Introduction}

\n Background Independence \cite{A64, A67, Giu06, ABook} 
is the modern form of the relational side of the Absolute versus Relational (Motion) Debate \cite{Newton, L, M, DoD-Buckets, ABook}.  

{            \begin{figure}[!ht]
\centering
\includegraphics[width=0.8\textwidth]{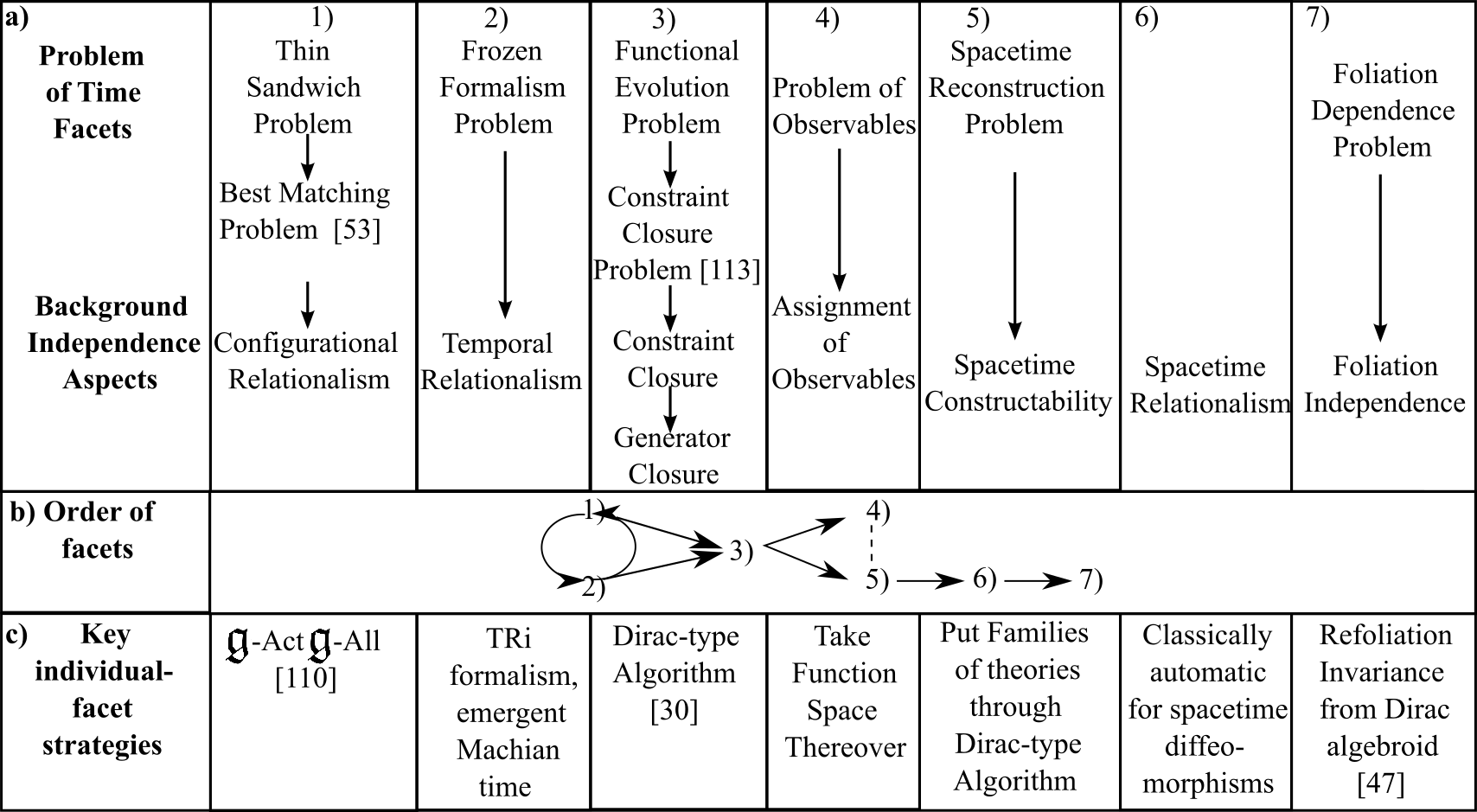}
\caption[Text der im Bilderverzeichnis auftaucht]{\footnotesize{a) Evolution of conceptualization and nomenclature from Kucha\v{r} and Isham's \cite{K92, I93} for local 
Problem of Time facets in the first row to the Author's Background Independence aspects \cite{APoT3, ABook} in the last. 
b) The order in which the facets are incorporated, itself resolving a longstanding problem \cite{K92, I93, K93}.
c) The key strategies used in the current Article's local resolution.                                             } }
\label{Evol-Fac}\end{figure}            }

\n Difficulites with implementation of Background Independence constitute the Problem of Time \cite{Battelle, DeWitt67, K92, I93, APoT, APoT2, APoT3, ABook}
The Problem of Time has nine facets corresponding to nine aspects of Background Independence. 
This refers to Isham and Kucha\v{r}'s \cite{K92, I93} conceptual classification of Problem of Time facets, 
which the Author converted            \cite{APoT2, APoT3, ABook} to a conceptual classification of theory-independent classical-or-quantum Background Independence aspects 
as per Fig 1's nomenclature.

\end{titlepage}

\n The first eight are local, of which the first seven already occur at the classical level.
The most significant part of their setting is brackets algebra \cite{Dirac, HTBook}, 
the first five being for canonical constrained theory, ending with recovering spacetime on such premises \cite{BFO, AM13}. 
One then switches to spacetime's own brackets algebra \cite{APoT3} and Foliation Independence \cite{T73, I93, TRiFol, ABook}.
Aspect 8 concerns nonuniqueness, in a specifically-quantum sense, i.e.\ solving the Multiple Choice Problems of Time \cite{K92, I93, Gotay00}.
Aspect 9 is rendering aspects 1-8) globally sound, i.e.\ solving the Global Problems of Time \cite{K92, I93, Epi-B}.
The Author recently gave \cite{A-Lett, ABook} a local resolution of the Problem of Time, i.e.\ of facets 1 to 7 
(in arenas for which these moreover form a consistent subset of the facets). 
This is the first resolution of its kind, strategies for individual facets having previously resisted joint treatment:  
the notorious `facet interferences' documented in \cite{K92, I93, ABook}.

\m

\n This topic presently appearing in the third Article on a series on Shape Theory, the first thing to note is the connection: 
that Shape(-and-Scale) Theory, alias {\it Relational Theory}, confers mastery of Aspect 2: Configurational Relationalism. 
Relational Theory can moreover be extended by placing a mechanics over the corresponding {\it relational spaces} -- shape(-and-scale) spaces -- 
which now indeed play the role of configuration spaces for such mechanics. 
These are known as {\it Relational Mechanics}: Shape(-and-Scale) Mechanics.
These provide a working theory \cite{BB82, B03, FORD, FileR, AObs2, AObs3, ABook} of {\sl all four} of aspects 1 to 4 of Background Independence.  
It was this move which, firstly, provided many of the conceptual insights into what aspects of Background Independence underlie Isham and Kucha\v{r}'s Problem of Time facets. 
Secondly, it resolved the problem of `in which order' one is to approach the facets. 
Thirdly, it gave how to jointly mathematize aspects 1 to 7 so as to exorcise `facet interferences'. 

\m 

\n A first message to Shape Theorists working in Statistics and/or Applied Geometry is thus as follows. 
{\sl your subject has recently played a major part in resolving a 50-year-old fundamental problem of Theoretical Physics, 
which is moreover modern Physics' version of Leibniz' and Mach's relational objections to Newton}.  
These considerations follow from building a Mechanics, rather than a Statistics, on Kendall's shape(-and-scale) spaces \cite{Kendall84, Kendall}.  
This Shape Statistics \cite{Kendall, Bhatta, DM16, PE16} also being greatly innovative and widespreadly applicable throughout the STEM subjects,   
the underlying {\sl Shape Geometry and Shape Topology are very promising subjects of study indeed}.
Such Relational Mechanics is outlined in Sec 2 alongside being used as an example demonstrating the nature of, and strategies for, 
implementing each of the first four aspects of Background Independence in this setting.  

\m 

\n For the remaining three local aspects of Background Independence to be present, 
one needs a theoretical framework that is general enough to be capable of accommodating relativistic theories. 
To this end, we introduce the dynamical formulation of General Relativity (GR) \cite{ADM} in Sec 3, 
giving its analogues of shape(-and-scale) spaces and its versions of the first four facets and underlying aspects 
\cite{DiracObs, BSW, Dirac, DeWitt67, Battelle, HTBook, K92, I93, BF, B94I, BFO, ABeables, TRiPoD, AObs2, AObs3, DO-1, ABook}. 
In Sec 3.9, we then show how rigidity permits \cite{BFO, AM13, ABook} recovery of local SR spacetime and a key feature of GR's dynamics from a general spatial ansatz. 
This implements the fifth aspect: Spacetime Construction.
This resolved, we handle spacetime's own Relationalism \cite{APoT3, ABook} in Sec 3.10 and Foliation Independence \cite{T73, I93, TRiFol, ABook} in Sec 3.11.  
We next outline the quantum version of the Problem of Time in Sec 4.  

\m

\n We subsequently entertain the global classical Problem of Time in Sec 5. 
\n Reduced spaces are key to how 2) is handled once and for all at the classical level in our local resolution of the Problem of Time.
For spaces with symmetry, these are usually stratified manifolds as per Article II; the 1- and 2-$d$ similarity and Euclidean cases are, exceptionally, manifolds.  
Stratified manifolds arising from quotienting are moreover prone to having merely-Kolmogorov, rather than Hausdorff, separation as per Article II again.
Furthermore elsewhere-habitual excision (and more occasional unfolding) methods for handling these are inapplicable in the physical context.  
Reduced spaces being in general stratified manifolds rather than manifolds 
is indeed one of the Global Problems \cite{Epi-B} encountered in attempting to extend our local resolution of the Problem of Time to a global resolution. 
This is moreover exacerbated by mere Kolmogorovness raising its head widespreadly (within the models with ab-initio symmetric absolute spaces).
\n Article I's alternative of working with non-symmetric absolute space is also contemplated in Sec 5.4 from a Background Independence point of view.  
Two other upcoming Articles \cite{Forth} will introduce furtherly `out of the box' alternatives at this critical juncture. 
Some further global issues \cite{Epi-B} with the Problem of Time are also outlined (zeros, domain of validity of approximations, constraints, observables and patchings). 

\m

\n This is furthermore consistent as a multiple-localities venture within a global treatment by its reduced spaces being Hausdorff paracompact (HP), which admit a Shrinking Lemma (Sec 6).
The corresponding Wheeler's superspace of GR is moreover also HP.
Among $N$-point-particle theories, then, Article II's HP reduced spaces guarantee selects a very small subset of theories that are concretely protected. 

\m 

\n Note moreover (Sec 7) that our Relational Mechanics implementing aspects 1 to 4 refers in particular to                 Kendall's similarity group model 
                                                                                                        and the Leibnizian setting's Euclidean group model.
Aside from these models' local resolution of the Problem of Time having been worked out in detail, especially in 1- and 2-$d$ (\cite{FileR, APoT2, APoT3, ABook}, 
considering a local resolution in the first place is particularly justified (Sec 5) by this case's global features: a point first unveiled in the current Article. 

\m  

\n All in all, while 1-4), 8), 9) can be posed \cite{Epi-C} for all theories [and 1-9) for all relativistic theories], 
meaning level-by-level in mathematical structure: affine, projective, conformal, topological manifold, topological space..., which can be resolved is on a case-by-case basis. 
In particular, our HP space arguments preclude affine and projective shape mechanics, and conceiving in terms of shapes in space is preferred over doing so in spacetime.
A substantial Selection Principle for {\it Comparative Background Independence} is thus born, this being a large and fundamental area of study.
It can indeed be argued \cite{ABook} that Comparative Background Independence is in the same ballpark as 
Background-Dependent Quantum Gravity as subfields of Background Independent Quantum Gravity.
This is thus of interest to Fundamental and Theoretical Physics, especially to `the GR side' of Quantum Gravity and classical dynamics of GR.  

\m  

\n{\bf Extension 1} The current Article has so far treated Background Independence at the flat and differential-geometric levels of structure.
Implications for Comparative Background Independence at all levels of structure (and so also topological manifold, topological space, metric space, point set...) 
are next outlined in Sec 7.5.
This further sharpens \cite{Epi-C}'s program.  

\m 

\n{\bf Extension 2} Our second message to Shape Theorists is as follows. {\it For you, the current Article is moreover an invitation.
Namely, that there has been a large recent expansion in Shape Geometry and Toplogy as motivated by Comparative Background Independence.
On the one hand, there is some overlap with models you already developed for Image Analysis and Computer Vision -- Affine and Projective Shape Theories.  
On the other hand, Comparative Background Independence includes plenty of other models as well.
E.g.\  conformal shapes, event shapes in spacetime, 
                         supersymmetric shapes, 
						 shape theory on any manifold rather than just on $\mathbb{R}^n$ or occasionally ${S}^n$ or $\mathbb{RP}^n$, 
for which the Articles I and II serve as an introduction.} 
Comparative Background Independence has a Background-Independent Statistics parallel as well.
By this, Topological Manifold and Topological Space Statistics \cite{NSW08, Bob-1, Ghrist, Bob-2} are a natural sequel to Shape Statistics.

\section{Background Independence}

\subsection{We use Lie's mathematics and in a brackets-centred setting} 

\n On the one hand, local resolution of the Problem of Time, perhaps unsurprisingly, largely consists of using Lie's differential-geometric local techniques.

\m 

\n 1) Lie derivatives \cite{Yano55, Stewart} 
\be 
\pounds_{\u{\sX}}
\ee 
with respect to some vector field $\u{\mX}$

\m 

\n 2) Lie brackets \cite{Stewart, FH}
\be 
\mbox{|[} \m \mbox{,} \, \m \mbox{]|}
\ee 
which are bilinear, antisymmetric and obey the Jacobi identity 
\be 
\mbox{|[} \, \mbox{|[} A \mbox{,} \, B \mbox{]|} \mbox{,} \, C \m \mbox{]|} + cycles = 0   \m . 
\ee 
\n 3) Lie algebras \cite{Serre-Lie, FH}: each Lie group's infinitesimal linear tangent spaces to the origin, whose extra product operation as an algebra in the Lie bracket.  

\m 

\n 4) Lie's `flow technique' for handling PDEs \cite{John, Lee2}.  
By this, a first-order quasilinear PDE 
\be 
\sum_{\alpha} a^{\alpha}(x^{\beta}, \phi) = b(x^{\beta}, \phi)
\ee 
for unknown variable $\phi$ is recast as a flow system of ODEs 
\be 
x^{\alpha \, \prime} = a^{\alpha}(x^{\beta}, \phi) \mma \phi^{\prime} = b(x^{\beta}, \phi) \m .  
\ee
$\prime := \pa/\pa\tau$ for some parameter $\tau$ along the flow.  

\m 

\n 5) Lie's integral approach to geometrical invariants \cite{Lie, G63, PE-1, DO-1} by use of such a flow technique. 

\m 

\n Among these structures, brackets algebras see particularly major and central use.

\m 

\n On the other hand, the {\sl arenas} these techniques are applied to are more modern than Lie's own setting.
This reflects that the Problem of Time's most severe form requires both QM and GR to be in play.  
These are 1920's and 1910's developments respectively, 
with GR's dynamics and canonical formalism moreover having to await the 1950's \cite{B52} and 60's \cite{ADM, Dirac}, 
whereas Lie's work is from the 1880's and 90's \cite{Lie}.
We comment further on the ensuing Lie-type mathematics beyond Lie's own remit in the Conclusion.

\subsection{Constraint providers}

\n The canonical brackets of constraints play a particularly significant role in Background Independence and the Problem of Time.  
A second dilemma moreover accompanies these. 

\m 

\n 1) {\bf Constraints are a Given Starting Point}, without question as to their origin.  
This position is quite common in Applied Mathematics.   

\m 

\n 2) {\bf Fundamental Principles act as Constraint Providers}.   
This position has arisen in Theoretical Physics, in particular due to the Wheeler school \cite{WheelerGRT, Battelle}.
Fundamental principles thus entertained can moreover have a philosophical as well as a physical character.
Indeed, producing constraints is a way of successfully sharply mathematically implementing certain philosophical principles.    
We introduce each aspect-and-facet piecemeal, before showing how a Large Method version of a Small Method for the piecemeal facet 
combines with all other local such to provide our local resolution of the Problem of Time.  
The first two facets thus introduced -- Configurational Relationalism and Temporal Relationalism -- 
are moreover both Constraint Providers that sharply mathematically implement philosophical first principles.

\subsection{Aspect 1: Configurational Relationalism}

\n {\it Configurations}  
\be 
\biQ
\ee 
are instantaneous snapshots of the state of a system. 
In the $N$ points-or-particles setting central to the current series of papers, we denote the incipient configurations by 
\be 
\biq
\ee 
with components $q^{iI}$ for $i$ a carrier space vector label and $I$ a point-or-particle label running from 1 to $N$.  

\m 

\n The space of all possible configurations for a given system $\FrS$ is the corresponding {\it configuration space}. 
\be 
\FrQ(\FrS)
\ee 
We now also remark upon the use of slanted font for finite-dimensional configurations and straight font for field configurations.  
We use mathfrak font for the corresponding configuration spaces (and more generally for spaces of objects), 
so as to immediately avoid confusion between objects of a given type and the space of all such objects. 

\m 

\n We additionally consider a group $\lFrg$ to act on our configurations, such that only those features invariant under this group are of relevance to our modelling situation.  
This amounts to working with further configuration spaces 
\be 
\w{\FrQ} \:=  \frac{\FrQ}{\lFrg} \m .    
\label{wQ}
\ee
The particular cases of this which we consider involve seeking to free oneself from a symmetric absolute space by quotienting out its automorphism group 
(to some level or other of geometrical structure). 
This addresses the relational horn \cite{BB82, B03, FileR, AMech, ABook} of the Absolute versus Relational (Motion) Debate \cite{Newton, L, M, DoD-Buckets}, 
often contemporarily also referred to as `seeking Background Independence' \cite{A64, A67, Giu06, ABook}.  
Article I moreover already provided many examples of this: 
shape(-and-scale) spaces for various $\lFrg$ acting on diverse manifold models $\FrM^d$ of carrier space $\FrC^d$.  
This modelling-irrelevance of the effect of $\lFrg$'s transformations -- so central to Shape(-and-Scale) Theory is moreover the basis of one of Background Independences's aspects, and, 
upon difficulties arising with its implementation, of one of the Problem of Time's facets. 
This Background Independence aspect's true-name is {\bf Configurational Relationalism}. 
It is moreover the {\sl first} aspect to consider in building our local Problem of Time resolution, by which this resolution can be said to have grown from Shape Theory...

\m 

\n There are two kinds of mathematical implementation for Configurational Relationalism.

\m 

\n 1) {\bf Direct implementation} \cite{FORD} in which one works solely with $\lFrg$-invariant objects $O(\w{\biQ})$.  

\m 

\n Knowledge of (\ref{wQ})'s geometry, and this geometry having the good fortune of being simple and/or already well studied, may be permissive of a direct approach.
Much of Kendall's work \cite{Kendall} on Similarity Shape Theory benefits from this being the case in 2-$d$ (and in 1-$d$, though this was already known previously, 
so Kendall concentrates on $\geq 2$-$d$).
In many cases, however, working directly is not possible because (\ref{wQ})'s geometry is unknown, complicated, or (e.g.\ as per Article II) highly pathological.  
There is however another approach which, at least formally, is {\sl completely general}. 

\m 

\n Such seeking can be either  indirect by applying a `Best Matching'               \cite{BB82, B03}     group action on unreduced configuration spaces 
                       or     by direct formulation on reduced configuration spaces \cite{FORD, FileR}: `relational spaces'.  
Kendall's Shape Theory \cite{Kendall84, Kendall89, Kendall} (see also \cite{Small, JM00, Bhatta, DM16, PE16} for reviews)
is a trove of reduced configuration space geometry work greatly advancing Relational Mechanics (see \cite{FileR, ABook, I, II, III, IV-2, Minimal-N}). 
Both approaches are additionally amenable to removal of absolute time as well -- Temporal Relationalism -- \cite{BB82, B94I, ABook}; 
there is also a Spacetime Relationalism \cite{APoT3, ABook, A-Lett} `freeing from absolute spacetime' counterpart.  

\m 

\n 2) {\bf Indirect implementation} \cite{FileR, APoT3, ABook}. 
First consider objects $O(\biQ)$.  
Next act on these with $\lFrg$, 
\be 
\s{\rightarrow}{\lFrg}_g O(\biQ)  \m .
\ee  
Finally, use some operation $\mbox{\Large S}$ over all of $\lFrg$ to eliminate the object's $\lFrg$-dependence: 
\be 
O_{\nFrg\mbox{-}\si\sn\sv}  \:=  \mbox{\Large S}_{\sbig \m \in \m \nFrg} \s{\rightarrow}{\lFrg}_g O(\biQ) \m . 
\ee 
Let us refer to this method by its true-name, of {\bf $\lFrg$-act $\lFrg$-all Method}.   

\m 

\n The most elementary and familiar example of this is {\sl group averaging}, 
which features in undergraduate-level Group Theory and Representation Theory courses.\footnote{Cauchy \cite{Cauchy} is the first known proponent;  
this entered common knowledge among mathematicians with Burnside's work \cite{Burnside} (see e.g.\ \cite{Serre, FH, Bala-Combi} for basic modern use.
This moreover, to the Author's best knowledge, makes Cauchy the first user of any kind of  $\nFrg$-act $\nFrg$-all Method}
%
Here the `all' operation is group averaging, 
\be 
\mbox{\Large S} = \frac{1}{|\lFrg|}\sum_{\sbig \m \in \m  \nFrg} \m \mbox{ (finite groups) or } \m  
\mbox{\Large S} = \frac{1}{\int_{\sbig \m \in \m \nFrg}\d \bigg_{\sH\sa\sa\sr}  } \, \int_{\sbig \m \in \m \nFrg} \d \bigg_{\sH\sa\sa\sr} \cdot \m \mbox{ (compact groups)}   \m .  
\ee 
Other possible `all' operations are, firstly, forming the group sum or group integral as appropriate.
Among the many possible such, let us first remark upon {\it Kendall's comparer} between similarity shapes
\be
\mbox{(Kendall $\lFrg$-Dist)}  \es  \mbox{min}_{\u{\theta} \m \in \m SO(d)} \, \u{n} \, \u{\u{Rot}}_{\u{\theta}} \, \u{n} \m   
\label{Kend-2}
\ee
for normalized centre-of-mass frame preshape vectors $\u{n}$. 
Secondly, on taking the sup or inf over the group, whether combinatorially or by appropriate Calculus.
The latter includes in particular extremization of action principles in accordance with the Calculus of Variations. 
One can moreover generalize from $\FrQ$ dependence to such as tangent space $\FrT(\FrQ)$ dependence or cotangent space $\FrT^*(\FrQ)$ dependence.
By this, configurations-and-velocities or configurations-and-momenta dependent cases can be included.  
It is also worth mentioning the great generalization by the insertion of an arbitrary sequence of maps between the `act' and `all' operations:  
\be 
O_{\lFrg\mbox{-}\si\sn\sv}  \:=  \mbox{\Large S}_{\sbig \m \in \m \nFrg} \circ Maps \m \circ \s{\rightarrow}{\nFrg} O_{\sbig} (\biQ) \m . 
\ee 
By this, e.g.\  Principles of Dynamics actions can feature among the entities that can be extremalized to produce $\lFrg$-invariant objects.

\m 

\n So let us construct a Mechanics \cite{BB82, B03, FileR, AMech} indirectly on shape(-and-scale) space.
We firstly $\lFrg$-Lie-drag-correct the action's velocities according to 
\be 
\dot{\u{q}}^I \m \longrightarrow \m \dot{\u{q}}^I - \pounds_{\sbig}\dot{\u{q}}^I
\label{Lie-Drag-1}
\ee 
The explicit form taken in the $Eucl(d)$ case \cite{BB82} is 
\be 
\dot{\u{q}}^I \m \longrightarrow \m \dot{\u{q}}^I - \u{a} - \u{b} \cr \u{q}^I \m ; 
\ee 
subtract off $c \, \uq^I$ as well to obtain the $Sim(d)$ case \cite{B03}.  
For now we can consider this for a standard Euler--Lagrange action input, with our corrections introducing Lagrange multiplier $\lFrg$-auxiliary coordinates. 

\m 

\n We secondly obtain the equations arising from variation of our action with respect to these $\lFrg$ auxiliary variables; these are Lagrange multiplier equations.  
These are constraints, by which Configurational Relationalism is a Constraint Provider. 
These constraints are moreover first-class and linear in their momenta, by which we denote them by 
\be 
\bFlin = 0 \m , 
\ee 
indexed as $\Flin_L$ for $L$ running over all of a theory's first-class linear constraints.
It is also hoped, if not yet confirmed, that these are $\lFrg$-gauge constraints $\bGauge$.  
By these constraints arising, our resolving $\lFrg$-invariance by incorporating {\sl further} $\lFrg$-dependent variables into our formalism is vindicated. 
For constraints of this kind \cite{Dirac, HTBook} use up two degrees of freedom per $\lFrg$ generator, 
thus taking into account bother the $\lFrg$ auxiliary's $\lFrg$-dependence and that of the original uncorrected action. 

\m 

\n We thirdly complete extremalization over $\lFrg$ by solving the Lagrangian variables $(\biQ, \dot{\biQ}, \bigg)$ form of these constraints for the $\lFrg$ auxiliaries themselves. 
We call these solutions `best-matched' and our three-step procedure above {\it Best Matching} or {\it the Barbour comparer} \cite{BB82, B94I}.  
We can moreover substitute these `best-matched' solutions back in the original action to obtain a reduced action, which amounts to completing Lagrange multiplier elimination.  
Henceforth, we can proceed via this reduced action. 

\m 

\n In the few fortunate cases admitting a direct formulation, we can write down a Mechanics action directly on $\w{\FrQ}$ \cite{FORD, FileR}.  
We call this a relational action; \cite{FileR} moreover showed that reduced and relational coincide for sim and Eucl relational theories, 
by which the joint-summary name {\sl r-formulation} is appropriate in cases of confluence.  

\m 

\n Finally, Principles of Dynamics actions are scarcely the only place we will need to apply this method in dealing with Background Independence. 
Other instances are forming $\lFrg$-independent notions of distance, information, correlation and quantum operator. 

\m 

\n This gives our first distinction between a Small Method: Group Averaging or Best Matching, and a Large Method: $\lFrg$-act, $\lFrg$-all. 
On the one hand, Small Methods deal with a single facet piecemeal, but very generally fail to deal even with that same facet in the presence of further facets. 
On the other hand, Large Methods have whatever generality it takes to successfully combine the facet in question with all other facets.

\subsection{Aspect 2: Temporal Relationalism}

\n The familiar Euler--Lagrange action 
\be 
{\cal S}  \es  \int L \, \d t^{\sN\se\sw\st\so\sn} 
          \es  \int \{T - V\} \d t^{\sN\se\sw\st\so\sn} 
          \es  \int \left\{ \frac{1}{2} M_{AB} \frac{\d Q^A}{\d t^{\sN\se\sw\st\so\sn}} \frac{\d Q^B}{\d t^{\sN\se\sw\st\so\sn}} - V(\biQ) \right\}\d t^{\sN\se\sw\st\so\sn}
\ee 
makes reference to Newtonian time $t^{\sN\se\sw\st\so\sn}$.  
The other substructures in use here are the {\it potential term} 
\be 
V = V(\biQ) \m , 
\ee
and the kinetic term 
\be 
T  \es  \frac{1}{2} M_{AB}(\biQ) \frac{\d Q^A}{\d t^{\sN\se\sw\st\so\sn}} \frac{\d Q^B}{\d t^{\sN\se\sw\st\so\sn}}   \m ,  
\ee 
where $M_{AB}(\biQ)$ is the kinetic metric on the $\biQ$'s configuration space $\FrQ$.  
The Lagrangian variables of this formulation are
\be 
\biQ \mma \dot{\biQ} \mma \dot{\m} := {\d }/{\d t^{\sN\se\sw\st\so\sn}}                                                    \m .  
\ee 
Since $t^{\sN\se\sw\st\so\sn}$ is extraneous, we replace this formulation with the {\it Jacobi action}. 
A first form of this is 
\be 
{\cal S}  \es  \sqrt{2} \int \sqrt{E - V} \sqrt{M_{AB} \frac{\d Q^A}{\d \lambda} \frac{\d Q^B}{\d \lambda}} \, \d \lambda  
          \es  \     2 \int \sqrt{W \, T_{\lambda}} \d \lambda                                                                                                               \m , 
\ee 
where velocities and integration are with respect to a `label-time' parameter $\lambda$; 
it is then these velocities which enter the kinetic term $T_{\lambda}$'s elsewise-standard definition.   
This is moreover {\it Manifestly Reparametrization Invariant}, since switching to a label-time $\mu$ gives an equivalent action by cancellation of the label-time coordinate changes, 
$$ 
\sqrt{E - V} \sqrt{M_{AB} \frac{\d Q^A}{\d \lambda} \frac{\d Q^B}{\d \lambda}} \, \d \lambda                                                        \es
\sqrt{M_{AB} \frac{\d Q^A}{\d \mu}\frac{\d \mu}{\d \lambda} \frac{\d Q^B}{\d \mu}\frac{\d \mu}{\d \lambda}} \, \frac{\d \lambda}{\d \mu} \, \d \mu  \es 
$$
\be 
\sqrt{M_{AB} \frac{\d Q^A}{\d \mu}\frac{\d Q^B}{\d \mu}}  \frac{\d \mu}{\d \lambda} \frac{\d \lambda}{\d \mu} \d \mu                                \es
\sqrt{E - V} \sqrt{M_{AB} \frac{\d Q^A}{\d \mu} \frac{\d Q^B}{\d \mu}} \, \d \mu                                                                                             \m . 
\ee 
One can furthermore rewrite this action in {\it Manifestly Parametrization Irrelevant} form  
\be 
{\cal S}  \es  \sqrt{2} \int \sqrt{E - V(\biQ)}  \sqrt{M_{AB}(\biQ) \d Q^A \d Q^B}  
          \es  \sqrt{2} \int \sqrt{W} \,         \d s  
          \es  \int                     \d J                                                                                                                                 \m , 
\ee 
and reconceive of this as a {\it geometrical action}\footnote{Our action is commonplace 
in the Dynamics \cite{Arnold} and Celestial Mechanics literatures in precisely this dual geometrical action conception.}
on the corresponding configuration space geometry, thus never making any mention of meaningless parameters in the first place. 
This reconception is in terms of change 
\be 
\d \biq 
\ee 
in place of velocity, of kinetic arc element 
\be 
\d s = \sqrt{M_{AB} \d Q^A \d Q^B}
\ee 
in place of kinetic energy, and of Jacobi arc element 
\be 
\d J = \sqrt{ 2 W} \, \d s =  \sqrt{2 \{ E - V \} }\sqrt{M_{AB} \d Q^A \d Q^B}
\ee 
in place of the Lagrangian.
\be 
W = W(\biQ) := E - V(\biQ)
\ee 
is the {\it potential factor}, where $E$ is the total energy of the system.  

\m 

\n In this way, Lagrangian variables are replaced successively by 
\be 
\biQ \mma \biQ^{\prime} \mma \mbox{}^{\prime} := {\d }/{\d \lambda}
\ee 
and 
\be 
\biQ \mma \d\biQ  \m : 
\label{Q-dQ}
\ee 
{\it geometrical mechanics} alias {\it Jacobi--Mach variables}.  

\m 

\n The premise of a geometrical action itself is to view dynamics as a geodesic on the corresponding configuration space geometry. 

\m

\n In summary, our guiding principle is the following. 

\m 

\n{\bf Leibnizian Temporal Relationalism} \cite{L} There is no time at the primary level for the universe as a whole. 

\m 

\n{\bf `Jacobi--Mach' implementation of Temporal Relationalism} \cite{APoT2, ABook} Make no reference to extraneous times (such as Newton's), nor to label times.  

\m 

\n {\sl Ab initio}, this leaves us working with Jacobi's action in terms of Jacobi--Mach variables. 
For the above fundamentally-motivated range of theories, moreover, the Jacobi action is physically equivalent to the Euler--Lagrange one. 
In its first form, this follows by, firstly, parametrizing $t^{\sN\se\sw\st\so\sn}$ in terms of $\lambda$.
Secondly, noting that the subsequent action depends on ${\d t^{\sN\se\sw\st\so\sn}}/{\d \lambda}$ alone rather than on $t^{\sN\se\sw\st\so\sn}$ itself, 
so that Routhian reduction is applicable.  
Performing this, and identifying the constant introduced as the total energy $E$, one arrives at the Jacobi action.  

\m 

\n Jacobi's principle works as follows. 
Dirac \cite{Dirac} proved that reparametrization-invariant actions must encode at least one primary constraint. 
Barbour \cite{BB82, B94I} showed that, for quadratic Jacobi actions, a `direction-cosines' or `Pythagorean' working gives this constraint to be quadratic as well: 
\be 
\scE  \:=  \frac{1}{2} \, N^{AB} P_A P_B + V(\biQ)  
        =  E                                          \m , 
\ee 
an equation known usually as `constant energy constraint', but to which Barbour attributed a different meaning in the current context. 
$N^{AB}$ is here the inverse of the kinetic metric $M_{AB}$, and $P_A$ are the momenta
The Author \cite{FileR, ABook} proved that all of these arguments transcend to parametrization-irrelevant actions and to geometric actions dual thereto. 
This necessitates rewriting the Principles of Dynamics formula for momenta: 
\be 
P_A \:= \frac{\pa L}{\pa \dot{Q}^A}                  \m \mbox{ to } \m 
P_A \:= \frac{\pa L_{\lambda}}{\pa Q^{A \, \prime}}  \m \mbox{ to } \m  
P_A \:= \frac{\pa \, \d J }{\pa \, \d Q^A}                 \m , 
\label{Mom-Form}
\ee 
where dot is derivative with respect to $t^{\sN\se\sw\st\so\sn}$, and dash is derivative with respect to $\lambda$.
The first two are standard, whereas the third is equivalent to the second by the `cancellation of the dots' Lemma.  

\m 

\n Barbour's interpretation of this quadratic constraint is moreover as {\it an equation of time}. 
For it can be rearranged to give a formula for a notion of time with respect to which motion is simple (a desirable property of timefunctions, as emphasized e.g.\ in \cite{MTW}):
\be 
t^{\se\sm} = \int \frac{\d s}{\sqrt{2 \, W}} = \int \frac{ \sqrt{M_{AB}(\biQ) \d Q^A \d Q^B}    }{\sqrt{2 \{E - V\}}}  \m . 
\label{tem}
\ee 
Because of this, I term this type of constraint a {\it Chronos constraint}, denoting all examples of such by 
\be 
\Chronos  =  0  \m . 
\ee 
Keeping track of which objects in a Background Independence scheme alias Problem of Time resolution is moreover an issue. 
For this requires many objects, many of which are new to address aspects or facets and/or newly formulated or newly interpreted because traditional versions of them 
succumb to facet interference \cite{K92, I93}.  
Because of this, we jointly denote constraints -- a sizeable subclass of such objects -- by the undersized calligraphic font, 
so that constraint status can immediately be read off the formalism.  
Full-sized calligraphic font is already in play as notation for functionals, as already evidenced by the Principles of Dynamics actions ${\cal S}$.  

\m 

\n $t^{\se\sm}$ denotes {\it emergent Machian time}.  
This name is justified by Mach's `time is to be abstracted from change' resolution of how an ab initio timeless relational world like Leibniz's 
                                                                                          comes to possess an emergent notion of time. 
More explicitly, we first implement Leibniz's ab initio timelessness by Jacobi's action principle's freedom from extraneous time and label-time reparametrization invariance. 
The latter feature can moreover be recast, via parametrization irrelevance, as a geometrical action with no reference to a label-time parameter at all. 
Then my uplift of Dirac's argument to apply to such geometrical actions enforces as a primary constraint 
                                                                                    a quadratic constraint 
																					that is to be viewed as an equation of time $\Chronos$. 
Working with the geometrical action, this equation moreover gives an emergent timefunction as an explicit {\sl functional} of change, 
by which it constitutes a mathematical implementation of `time is to be abstracted from change' and thus merits its Machian monicker.
This is moreover only possible if we have passed from reparametrization invariant actions to parametrization irrelevant-or-dually-geometrical actions. 
I.e.\ to actions in terms of configuration-and-change variables (\ref{Q-dQ}), which thus justify the Machian half of their `Jacobi--Mach' sobriquet.

\m 

\n In more detail, while any \cite{R04} and all \cite{B94I} change have been used elsewhere, 
what a detailed analysis reveals to actually be required (Chapter 15 of \cite{ABook}) is a {\it sufficient totality of locally relevant change} (STLRC). 
The time abstracted from this is {\it generalized local ephemeris time} (GLET). 
This is even {\sl even more like} ephemeris time \cite{Clemence} that \cite{B94I}'s notion of time 
(inspired by a mixture of \cite{Clemence} and further Leibnizian features, which I, however jettisoned as operationally unsuitable for physical practice).  

\m 

\n For Temporally-Relational mechanics, moreover, the $t^{\se\sm}$ thus obtained {\sl mathematically coincides} with $t^{\sN\se\sw\st\so\sn}$, 
while now resting on relationally-acceptable foundations and possessing an emergent character. 

\m 

\n This approach can be surmised by 
\be 
\mbox{\sl taking Jacobi's action principle seriously enough to rederive the rest of Physics concordantly}. 
\label{Jac-Ser}
\ee 
We do so to have guaranteed implementation of Temporal Relationalism.
This serves, firstly, to cast Physics into a Leibnizian relational form from which emergent Machian time emerges. 
Secondly, to remove once and for all all facet interferences that partially involve the Temporal Relationalism aspect (previously known as `Frozen Formalism Problem facet').

\m 

\n (\ref{Jac-Ser}) consists of the following. 
The Jacobi action is but a starting point for reformulating the Principles of Dynamics (PoD), and the rest of Physics, in Temporal Relationalism implementing (TRi) terms. 
This requires replacing `around half' of the structures we use: {\it TRiPoD} \cite{TRiPoD}, 
                                                                {\it TRiFol} (spacetime foliations) \cite{TRiFol}, TRiCQT (Canonical Quantum Theory), TRiPIQT (Path Integral Quantum Theory). 
The `other half' turns out to be already-TRi in its standard formulation.  
See \cite{ABook} for a full account of these reformulations, though some examples of them are given in Secs 2, 3 and 4 of the current Article,  
and these already suffice to indicate, moreover, that 'much of the PoD that leads naturally to QM is already-TRi' (see Sec 8.1 for details). 
Among the objects encountered so far, configurations $\biQ$ are already TRi-invariant. 
While momentum {\sl formulae} (\ref{Mom-Form}) need TRi-updating, the {\sl notion of momentum itself} is already TRi-invariant as well: 
a further superiority of momentum over velocity as a concept and progenitor of further concepts and formulations. 

\m 

\n This further illustrates the difference between, on the one hand, a Small Method for dealing with one Problem of Time facet: here use of Jacobi actions.
And, on the other hand, a Large Method for dealing with all of that facet's interferences with further facets: here the recasting of all of Physics in TRi form. 
This is what it takes to {\sl remain entirely} within Temporal Relationalism as we handle each subsequent facet in turn.

\subsection{Configurational Relationalism tailored to also fit Temporal Relationalism}

\n Let us firstly consider $\lFrg$-Lie-drag-correcting our first Jacobi action's label-velocities \cite{BB82, B03}. 
This however spoils its Manifest Reparametrization Invariance property.  

\m 

\n We get round this first facet interference by using instead label-velocity representations of the generators \cite{ABFO}. 
This succeeds in preserving Manifest Reparametrization Invariance, but shifts the auxiliary $\lFrg$-variables' Principles of Dynamics status from multipliers to cyclic coordinates. 
None the less, {\it free end point value variation} \cite{FEPI, FileR} ensures the outcome of varying with respect to the $\lFrg$-variables transcends this reformulation.  
Let us also identify this `best-matched' solution followed by substituting back into the original action as a Routhian Reduction \cite{Lanczos}. 

\m  

\n To work with the geometrical action, moreover, we require \cite{FileR} to Lie-drag-correct with $\lFrg$-auxiliaries that are themselves changes, 
\be 
\d{\u{q}}^I \m \longrightarrow \m \d{\u{q}}^I - \pounds_{\d \sbg}{\u{q}}^I
\label{Lie-Drag-2}
\ee    
Free end point value variation carries over. 
Our procedure now moreover involves not Routhian reduction but a TRi-variant thereof: {\it dRouthian reduction} \cite{TRiPoD, AM13}. 
Namely, elimination from the original action of the properly-TRi Jacobi--Mach variables formulation's cyclic differentials 
                                                                                       rather than of cyclic velocities in the usual Routhian reduction.
Using this final method, we succeed in jointly incorporating Temporal and Configurational Relationalism at the level of the action.  

\m 

\n In greater generality, we need to adjust the $\lFrg$-act $\lFrg$-all Method's $\lFrg$ auxiliaries to be TRi-represented 
to jointly incorporate Temporal and Configurational Relationalism. 

\m 

\n Note finally that we need to {\sl iterate} implementing Temporal and Configurational Relationalisms until consistency is attained.  
This is represented by the loop in Fig 1.b). 

\m 

\n A first reason for this is that finding an explicit $t^{\se\sm}$ in the presence of Lie-drag corrections requires prior elimination of the $\lFrg$-auxiliary variables. 
This is clear from the $d\biQ$ in its formula still carrying these corrections unless one substitutes in the best-matched values for these, or, alternatively, 
starts afresh with the reduced action after performing this elimination therein.  
The next subsection's facet's check provides a second reason.

\subsection{Aspect 3: Constraint Closure}

We next need to assess whether the constraints provided by each of Temporal and Configurational Relationalism are consistent.  

\m 

\n The {\it Dirac Algorithm} \cite{Dirac, HTBook} is a powerful Hamiltonian-variables-level tool for assessing whether a set of constraints in hand is, or can be made, consistent. 
Its end product is a {\it constraint algebraic structure} that is closed under Poisson brackets.  

\m  

\n The following good fortunes occur.
Hamiltonian variables are (\biQ, \biP), which are all already-TRi. 
Constraints in this formulation are just function(al)s of these variables, 
\be 
\sbcC(\biQ, \biP) = 0 
\ee 
and thus are already-TRi as well.  
Poisson brackets, being defined in terms of $(\biQ, \biP)$, are also already-TRi. 
Finally, Poisson brackets algebras of constraints are by extension already-TRi too. 
All of these observations moreover transcend to other relevant notions of constraints (e.g.\ quantum-level constraints), 
brackets (e.g.\ classical Dirac Brackets and quantum brackets), 
and constraints brackets algebraic structures. 

\m 

\n While some other parts of the Dirac Algorithm are not TRi, they can be readily modified to comply, giving the {\sl TRi almost-Dirac Algorithm} \cite{TRiPoD, AM13, MBook, ABook}.  
These modifications firstly comprise replacing the bare Hamiltonian $H$ with a differential Hamiltonian $\d H$ to match the Lagrangian $L$ 
being replaced by the Jacobi arc element $\d J$.
Secondly, the Legendre transformation from $L$    to   $H$     with its extra                
\be 
\biP \dot{\biQ}
\ee 
term is replaced by a TRi-Legendre transformation from $\d J$ to $\d H$ now with an extra Liouville form 
\be 
\biP\d \biQ
\ee 		
term instead.  
Thirdly, Dirac's total Hamiltonian (and similar) is a multiplier appending of constraints (the index $F$ runs over all first class constraints)
\be 
H_{\sT\so\st\sa\sll} = H_{\sb\sa\sr\se} + m_{F} \scC^F
\ee 
and thus TRi-violating, so it is replaced by my TRi {\it differential-almost-Hamiltonian}, 
\be 
\d A = \d H_{\sb\sa\sr\se} + \d c_F \scC^F  \m . 
\ee 
This is a {\sl differential-almost}-Hamiltonian since it can contain cyclic differentials of auxiliary quantities as well as configurations and momenta of 
{\it quantities of partial or total physical content}. 
Fourthly, this has a knock-on effect in requiring Poisson brackets to be replaced by mixed Peierls--Poisson brackets (see \cite{DeWittBook} for an account of Peierls brackets). 
This is however mitigated by the {\sl physical quantities} all obeying Poisson brackets relations, with the Peirels sector being relegated to an `auxiliary fluff' sector. 
Thus converted, we have my TRi Dirac-type Algorithm, whose physically relevant sector functions in very close analogy to the Dirac Algorithm itself, 
while now managing to be TRi and thus free of facet interferences with Temporal Relationalism.  

\m  

\n Now, given our two constraint providers, our Dirac-type Algorithm has three checks to make. 

\m 

\n 1) Whether Configurational Relationalism's $\bFlin$ is self-consistent 
(and a fortiori $\bGauge$, and a plus fortiori $\lFrg$-$\bGauge$ rather than being forced to extend to a larger group). 

\m 

\n 2) Whether Temporal Relationalism's $\Chronos$ is self-consistent. 

\m 

\n 3) Whether $\bFlin$ and $\Chronos$ are mutually consistent: the cross-brackets with one $\bFlin$ and one $\Chronos$ entry. 

\m 

\n Note that without 3), the Dirac Algorithm fails one's candidate theory in hand. 
Failure of 1) may be acceptable if one did not have ironclad reasons to want $\lFrg$ to be one's gauge group. 
Failure of 3) (say) would also be acceptable if it required some of the constraints from the other partition, 
here $\bFlin$, as integrabilities to ensure $\Chronos$'s algebraic closure. 

\m  

\n For $Sim(d)$ and $Eucl(d)$ Relational Mechanics, moreover, all of 1-3) succeed individually. 
In this way, we have succeeded in, firstly, obtaining a unified scheme for solving the first three Problem of Time facets. 
Secondly, in showing that the usual $Eucl(d)$-invariant relational $N$-body problem and the shape mechanics without scale based on Kendall's Similarity Shape Theory 
locally satisfy these three criteria. 
The former is boosted by the further observation that observables and rigidity consistently postcede this triple resolution, and can be approached in either order themselves. 
The latter, is boosted by Relational Mechanics having no notion of spacetime, and consequently no notion of Spacetime Relationalism, Spacetime Construction or spacetime 
Foliation Independence.   
By this, we are now only one facet away from a local resolution of the classical Problem of Time.

\m 

\n For subsequent use below, there is further interest in finding all the consistent subalgebras supported by the full constraint algebra. 
These form a bounded lattice $\lattice_{\sFrC}$ with the full constraint algebra as top element and the trivial algebra $id$ -- no constraints as generators -- as bottom element.
In particular, for Relational Mechanics, 1) and 3) succeeding individually means that both $\bFlin$ and $\Chronos$ are `middle' (non-extremal) elements in the lattice of constraint 
algebraic structures for these.
%
{            \begin{figure}[!ht]
\centering
\includegraphics[width=1.0\textwidth]{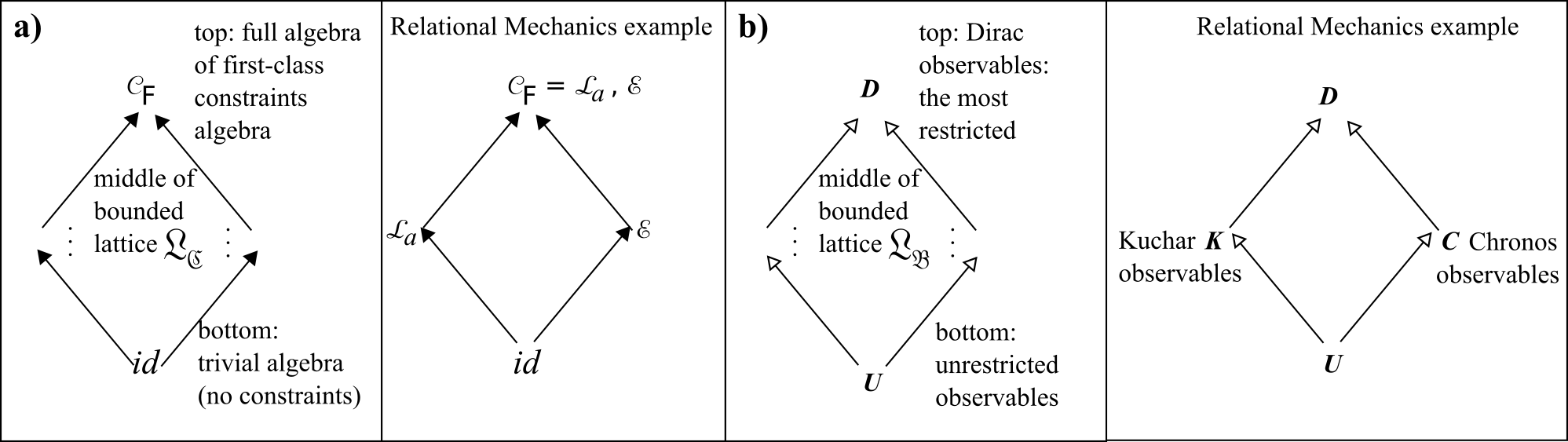}
\caption[Text der im Bilderverzeichnis auftaucht]{ \footnotesize{Lattice of subalgebras of a) constraints b) observables} }
\label{Const-Obs}\end{figure}            }

\subsection{Aspect 4: Expression in terms of Observables}

\n The most primary notion of observables involves {\it Taking a Function Space Thereover}, referring in the classical canonical case to over phase space
\be 
\Phase(\FrS)  \m : 
\ee 
the (\biQ, \biP) space as equipped with the Poisson bracket 
\be
\mbox{\bf \{} \m \mbox{\bf ,} \, \m \mbox{\bf \}}  \m .
\ee 
This notion's output are the {\it unrestricted observables}, 
\be 
\sbiU(\biQ, \biP)
\ee 
forming some function space of suitably-smooth functions, such as 
\be 
\FrU = \cC^{\infty}(\Phase(\FrS))  \m .
\ee 
\n Facet interference with the presence of constraints imposes the {\it commutation condition}  
\be 
\mbox{\bf \{}  \scC_A  \mbox{\bf ,} \, \sbiB \mbox{\bf \}}  \m  `='  \m 0  
\m \mbox{ i.e.} \es  
\left\{  
\s{\mbox{0} \m \mbox{ (strongly vanishing)}}{ {F_{AB}}^{B^{\prime}} \siB_{B^{\prime}}  }  
\right . 
\label{CO}
\ee 
on canonical observables functions, $\sbiB$. 
For this to be consistent \cite{ABeables}, the $\scC_A$ must moreover be a closed algebra of first-class constraints, by the Jacobi identity with $\sbcC$, $\sbcC$, $\sbiB$ entries. 
This {\sl decouples} finding canonical observables to 
{\sl after} establishing Constraint Closure of the constraints provided by Temporal and Configurational Relationalism, as delineated in Fig 1.b).  

\m 

\n Each closed subalgebra of constraints moreover furnishes its own notion of canonical observables.
The most salient two cases are no constraints, returning the unrestricted observables  $\sbiU$,      and all first-class constraints, 
                                               returning the {\it Dirac observables}   $\sbiD$ \cite{DiracObs}: the fully physical case.
All physical theories support these notions. 
Some physical theories support further intermediate notions of classical canonical observables. 
The totality of notions of observables form a bounded lattice $\lattice_{\sFrB}$ dual to that of constraint algebras, $\lattice_{\sFrC}$.  
Relational Mechanics supports both {\it Kucha\v{r} observables} $\sbiK$ -- commuting with the entirety of $\bFlin$ -- 
                               and {\it Chronos    observables} $\sbiC$:   commuting with                 $\Chronos = \scE$.
							   
\m

\n As regards not only defining suitable concepts of observables but also finding a full complement of observables,  
each type of observables itself forms an algebraic structure, by the $\sbiB, \sbiB, \sbcC$ Jacobi identity). 

\m 

\n As regards evaluating the observables of each kind, \cite{AObs2, ABook, PE-1, DO-1} further formulate the bracket conditions (\ref{CO}) as specific PDEs.
This is a subcase of Lie's integral method for finding invariant functions, uplifted moreover to a free characteristic problem for finding 
'suitably-smooth functions of phase space invariants' = observables.  

\m 
 
\n Let us end by ascertaining that Expression in terms of Observables is already-TRi, since Poisson brackets, constraints and constraint algebras are, 
as are the configurations and momenta that the new objects -- observables -- depend upon. 
Finally observables algebras and the lattice of subalgebras thereof are already-TRi as well.

\section{GR counterpart}

\n GR-as-Geometrodynamics moreover possesses a formulation along the lines of Shape Theory and Fig 1's multi-aspect account Background Independence more generally.

\subsection{Incipient configuration space $\Riem(\bupSigma)$} 

GR-as-Geometrodynamics refers to its most longstanding dynamical \cite{B52} and canonical \cite{ADM, WheelerGRT, Battelle, DeWitt67} formulation.  
The incipient configurations here are Riemannian 3-metrics $\bh$ with components $\mh_{ab}(\ux)$ (for $\ux$ coordinates in space) 
on a fixed 3-topology $\bupSigma$ interpreted as a spatial slice of GR spacetime, 
itself a semi-Riemannian 4-metric $\bg$ with components $\mg_{\mu\nu}$ on a 4-topology $\FrM$. 
These are analogous to Shape Theory's incipient $N-point$ configurations $\biq$ with components $q^{aI}$, 
with $\bupSigma$ playing a more loosely analogous role to Shape Theory's underlying carrier space $\FrC^d$.
As the 3-metric $\bh$ is a symmetric $3 \times 3$ matrix, it has 6 degrees of freedom per space point. 

\m 

\n The space formed by the totality of the $\mh_{ab}$ on a fixed $\bupSigma$ is GR's incipient configuration space 
\be 
\FrQ(\bupSigma) = \Riem(\bupSigma)                        \m  ; 
\ee
this is analogous to Shape Theory's constellation space, 
\be 
\FrQ(\FrC^d, N) = \bigtimes_{I = 1}^N \FrC^d  \m .
\label{Stel}
\ee 
We consider compact without boundary manifolds models for space $\bupSigma$, with $\mathbb{S}^3$ and $\mathbb{T}^3$ the most commonly considered specific spatial topologies.

\m 

\n $\Riem(\bupSigma)$ is supplied with its own metric by GR's action in split space-time form \cite{ADM}: the inverse DeWitt metric \cite{DeWitt67}, 
\be 
\mM^{abcd}  \:=  \sqrt{\mh} \{ \mh^{ac} \mh^{bd} - \mh^{ab} \mh^{cd}\}  \m : 
\ee   
the inverse of the DeWitt supermetric $\mN_{abcd}$
This is the analogue of the $\mathbb{R}^{Nd}$ Euclidean metric on Shape Theory's underlying configuration space, 
or more generally of the product metric on the constellation space (\ref{Stel}).
While the inverse DeWitt supermetric is indefinite, so is Event Shape Theory's incipient product space metric \cite{Event-Shapes}. 

\m 

\n Also note that for $\FrC^d = \mathbb{R}^d$; the analogy between relative Jacobi separation vectors $\u{\rho}^i$ and the $\mh_{ab}$, and so between relative space $\Frr(d, N)$, 
is comparable with the above one. 
We use this second analogy throughout the rest of this Article.

\subsection{Diffeomorphims $Diff(\bupSigma)$ in role of automorphisms} 

\n The spatial diffeomorphisms 
\be
\lFrg = Diff(\bupSigma)
\ee 
are GR's automorphism group acting on the incipient configurations that are regarded as physically meaningless.
These ara analogous to considering $Rot(d)$ acting on relative space, including in having scale, though it is now a {\sl local} notion of scale.

\m 

\n Working with the diffeomorphisms, is a differential-geometric level endeavour, implemented infinitesimally by the Lie derivative. 
In 3-$d$, diffeomorphisms use 3 degrees of freedom per space point.

\subsection{$\lSuperspace(\bupSigma)$ as a further relational space} 

\be
\Superspace(\bupSigma) \:=  \frac{\Riem(\bupSigma)}{Diff(\bupSigma)}   \m .  
\label{Intro-Superspace}
\ee
is Wheeler's {\it superspace} \cite{Battelle} first studied in detail by his contemporaries DeWitt \cite{DeWitt67, DeWitt70} and Fischer \cite{Fischer70}, 
and subsequently studied in further detail by Fischer and Moncrief \cite{FM96} and by Giulini \cite{Giu09}.

\m 

\n 3-Metric $\bf$ minus 3-diffeomorphism information leaves 3-geometric information $\bG^{(3)}$: 
a notion of scaled shape, analogous to Euclidean Relationalism's rotationally-invariant inner products. 
Counting this out, superspace has $6 - 3 = 3$ degrees of freedom per space point.

\subsection{$\CRiem(\bupSigma)$ as a further preshape space} 

\n We next consider $Conf(\bupSigma)$ of spatial conformal transformations as a further group of automorphisms with natural action on $\Riem(\bupSigma)$.  
This is analogous to considering $Dil$ acting on relative space.  
This has 1 degree of freedom per space point: a localized scale.  

\m 

\n Upon quotienting this out, the corresponding configurations are unit-determinant metrics,  
\be
\buu  \:=  \frac{\bh}{^3\sqrt{\mh}}
\ee
for $\mh := \mbox{det} \, \bh$.  

\m 

\n The space of these is {\it conformal Riem} 
\be 
\CRiem(\bupSigma)  \:=  \frac{\Riem(\bupSigma)}{Conf(\bupSigma)}    \m . 
\ee 
This has $6 - 1 = 5$ degrees of freedom per space point.
%
%
This is GR's counterpart of Kendall's preshape sphere, with the unit determinant metrics in place of unit-modulus vectors. 
This analogy extends to each being relatively simpler topologically and geometrically than the corresponding theory's other reduced configuration spaces.  
The supermetric on $\CRiem(\bupSigma)$ is moreover positive-definite.

\subsection{$\CS(\bupSigma)$ as a further shape space} 

\n Let us finally consider 
\be 
\lFrg = Conf(\bupSigma) \rtimes Diff(\bupSigma)
\label{CD}
\ee 
in the role of automorphism group; this is analogous to $Dil \times Rot(d)$ acting on relative space.  
This involves $4$ degrees of freedom per space point.

\m 

\n The corresponding reduced configuratons are now conformal 3-geometries $\bC^{(3)}$ \cite{Y74}.  

\m

\n{\bf Structure 5} The space of these is York's \cite{Y74} {\it conformal superspace} 
\be
\CS(\bupSigma) \:= \frac{\Riem(\bupSigma)}{Conf(\bupSigma) \rtimes Diff(\bupSigma)}    \m ,  
\ee 
-- the arena of the most successful GR initial-value problem approaches to date \cite{York} -- 
and which has since been further considerably studied by Fischer and Moncrief \cite{FM96}.
This is analogous to Kendall's shape space.

\subsection{Temporal and Configurational Relationalism for GR}
%
{            \begin{figure}[!ht]
\centering
\includegraphics[width=0.5\textwidth]{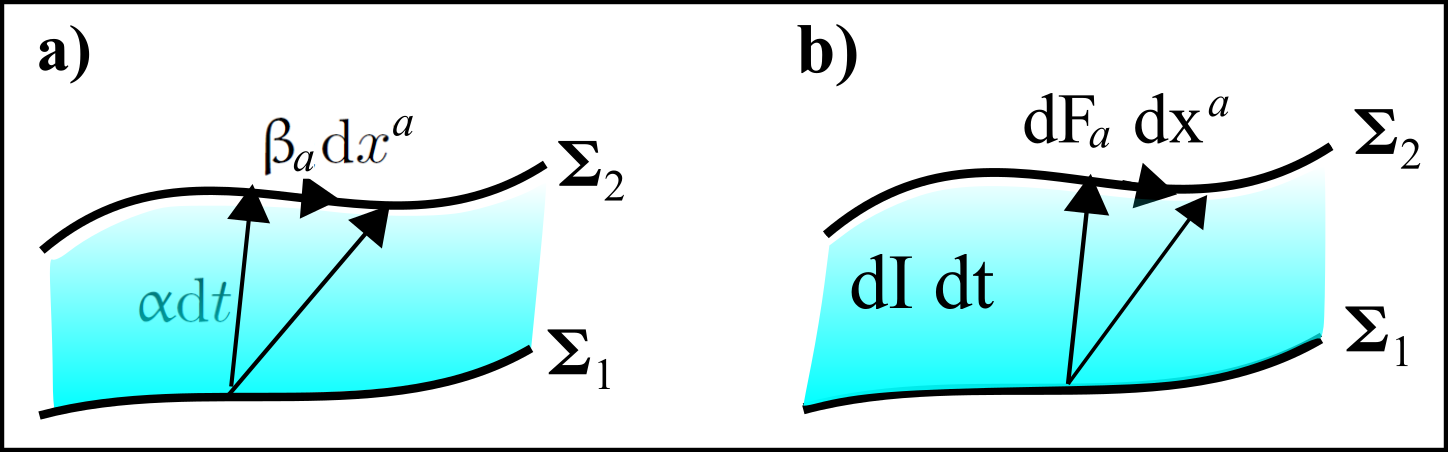}
\caption[Text der im Bilderverzeichnis auftaucht]{ \footnotesize{a) ADM kinematics \cite{ADM}. b) The A formulation's TRi kinematics \cite{FEPI}.
Furthermore, $\mD_a$ denotes spatial covariant derivative, ${\cal R}$ the spatial Ricci 3-scalar curvature, and $\slLambda$ the cosmological constant.} }
\label{Strutty}\end{figure}            }

\n Indirect formulation of Configurational Relationalism is necessary for GR. 

\m 

\n The Arnowitt--Deser--Misner (ADM) action for GR-as-Geometrodynamics is  
\be 
{\cal S}^{\sG\sR}_{\sA\sD\sM} \es \int \d\mt \int_{\sbSigma} \d^3x \sqrt{\mh} \, \upalpha 
\left\{
\frac{{\cal T}_{\sA\sD\sM}}{4\upalpha^2} +  {\cal R} - 2 \, \slLambda
\right\} \m , 
\ee
for GR kinetic term 
\beq
{\cal T}_{\sA\sD\sM} \es \mM^{abcd} 
\left\{ 
\frac{\pa \mh_{ab}}{\pa \mt}  - \pounds_{\u{\upbeta}} \mh_{ab} 
\right\}
\left\{ 
\frac{\pa \mh_{cd}}{\pa \mt}  - \pounds_{\u{\upbeta}} \mh_{cd}  
\right\}                                                            \m . 
\ee 
Consult Fig \ref{Strutty}.a) for the notation used; 
\be 
\mM^{abcd}  \:=  \sqrt{\mh} \{ \mh^{ac} \mh^{bd} - \mh^{ab} \mh^{cd} \}
\ee 
is the {\it inverse DeWitt supermetric} \cite{DeWitt67} alias {\it GR configuration space metric}.  

\m 

\n $\u{\upbeta}$ and $\upalpha$'s multiplier equations give, respectively, GR's {\it momentum constraint} 
\be 
\scM_a := - 2 \, \mD_b {\mp^b}_a = 0 
\ee 
and {\it Hamiltonian constraint} 
\be
\scH \:= \mN_{abcd} \mp^{ab} \mp^{cd} + \sqrt{\mh} \{ 2 \, \slLambda - {\cal R} \} \es 0   \m , 
\ee 
$\mp^{ab}$ is here GR's momenta -- a densitived and partly trace-removed version of the more widely familar extrinsic curvature tensor -- whereas 
\be 
\mN_{abcd} \:= \{\mh_{ac}\mh_{bd} - \mh_{ab}\mh_{cd}/2\}/\sqrt{\mh}  \m 
\ee 
is {\it DeWitt's supermetric itself} \cite{DeWitt67}. 
 
\m 

\n Temporal Relationalism is moreover now taken to eschew extraneous {\sl timelike} variables. 
By this, the ADM formalism will not do, since the lapse function it contains, signifying `time elapsed', is a such.  
We can however use the lapse's own multiplier equation to multiplier-eliminate the lapse out of the ADM action. 
In this way, we arrive at the Baierlein--Sharp--Wheeler action \cite{BSW}  
\be
{\cal S}^{\sG\sR}_{\sB\sS\sW} \es \int \d\lambda \int_{\sbupSigma} \sqrt{\mh} \sqrt{{\cal T}^{\sG\sR}_{\sB\sS\sW} \{{\cal R} - 2 \, \slLambda \}}  \m , 
\ee      
with kinetic term 
\be  
{\cal T}^{\sG\sR}_{\sB\sS\sW} \:= \mM^{abcd}
\left\{ 
\frac{\pa \mh_{ab}}{\pa \lambda}  - \pounds_{\u{\upbeta}} \mh_{ab} 
\right\}
\left\{ 
\frac{\pa \mh_{cd}}{\pa \lambda}  - \pounds_{\u{\upbeta}} \mh_{cd} 
\right\}
\ee 
now and henceforth using the undensitized version of $\mM^{abcd}$.  

\m  

\n Let us first note that this is a local comparer (in the sense of integrating after square-rooting).
In contrast, global comparers (square-root after integrating) can be found in \cite{DeWitt67} and \cite{BB82}; 
it took until \cite{B94I} for the local version to make it into the Relational Program.  

\m 

\n Let us next consider how, now that we longer have a lapse, the crucial Hamiltonian constraint $\scH$ arises from the local Baierlein--Sharp--Wheeler type actions. 
For these, $\scH$ is recovered as 1 primary constraint per space point by Dirac's argument about enforcing primary constraints.  
On the other hand, $\scM_a$ arises as before, from variation with respect to the ADM shift, $\upbeta^a$.

\m 

\n We next need to deal with the Configurational Relationalism so far breaking the Manifest Reparametrization Invariance, 
firstly by introducing cyclic velocities of the frame (Fig 3.b) as auxiliary variables in place of the shift.  
\n Its corrected kinetic term is  
\be 
{\cal T}^{\sG\sR}_{\sB\sF\sO\mbox{-}\sA} = \mM^{abcd}\left\{ 
\frac{\pa \mh_{ab}}{\pa \lambda}  - \pounds_{\pa\u{\sF}/\pa \lambda} \mh_{ab} 
\right\}
\left\{ 
\frac{\pa \mh_{cd}}{\pa \lambda}  - \pounds_{\pa\u{\sF}/\pa \lambda} \mh_{cd} 
\right\}  \m , 
\ee 
where $\u{\mF}$ is the TRi analogue of the ADM shift: the frame variable of Fig \ref{Strutty}.b).  
The corresponding action is \cite{FEPI}
\be 
{\cal S}^{\sG\sR}_{\sB\sF\sO\mbox{-}\sA}  \es  \int \d\lambda \int_{\sbupSigma}  \sqrt{\mh} \sqrt{{\cal T}^{\sG\sR}_{\sB\sF\sO\mbox{-}\sA} \{{\cal R} - 2 \, \slLambda \}}  \m .
\ee
\n This replacement of shift by velocity of the frame can be done at the level of the ADM action as well; 
if accompanied homogeneously by replacing the lapse by the velocity of the instant, $\mI$ (Fig \ref{Strutty}.b), we obtain an `A action' \cite{FEPI}.
Now passing from the A action to the BFO-A action \cite{ABook} is Routhian reduction in exact parallel to that from Euler--Lagrange to Jacobi actions for Mechanics. 

\m 

\n I subsequently gave a cyclic differential of the frame version of BSW-type action for GR \cite{PPSCT}; see \cite{FileR, ABook} for details.  
This is Manifestly Parametrization Irrelevant, or, dually, the geometrical action for GR.   
Its corrected line element is  
\be 
\d \scS^{\sG\sR}_{\sr\se\sll} = \mM^{abcd}\{\d  - \pounds_{\d \u{\sF}}\} \mh_{ab}\{\d - \pounds_{\d \u{\sF}}\}\mh_{cd}   \m  . 
\ee 
and the final relational action for GR is  
\be 
{\cal S}^{\sG\sR}_{\sr\se\sll} \es \iint_{\sbSigma}\d^{3}x \, \pa{\cal J}  
                               \es \iint_{\sbSigma}\d^{3}x   \sqrt{\mh}   \sqrt{ {\cal R} - 2 \, \slLambda} \, \d \scS^{\sG\sR}_{\sr\se\sll}    \m .
\label{S-Rel-d}  
\ee 
\n Now Configurational Relationalism and (now fully functional) Temporal Relationalism produce constraints $\scM_a$ and $\scH$ respectively. 

\m 

\n $\scH$, now viewed as a $\Chronos$ constraint, can furthermore be rearranged to give GR's emergent Machian time, 
\be
\mt^{\se\sm}_{\sG\sR}  \:=  \int_{\sbSigma} \frac{\d \scS_{\sG\sR}}{2\sqrt{ {\cal R} - 2 \, \slLambda}  } \m . 
\ee 
\n In the GR-as-Geometrodynamics context, Best Matching gives the Thin Sandwich.  
While this is `Machianized' by further development of actions, the PDE nature of the elimination remains invariant.  
PDE studies for this are provided in \cite{BO, BF}. 
`Thin Sandwich' was also the facet name in the early Problem of Time literature \cite{WheelerGRT}, retained through to the Isham--Kucha\v{r} classification.  
Barbour's Best Matching generalized the `Thin Sandwich' facet to a wide range of classical actions, 
and my $\lFrg$-act $\lFrg$-all approach for Configurational Relationalism to classical or quantum version to any level of structure for any theory.

\subsection{GR's Constraint Closure}

\n Let us first introduce two useful GR models as regards part of Background Independence and the Problem of Time that Relational Mechanics does not possess, 
alongside giving their constraint closure status. 

\m 

\n 1) Minisuperspace (spatially homogeneous GR) \cite{Magic}, which has a single finite constraint, and so closes trivially. 
 
\m  
 
\n 2) Slightly Inhomogeneous GR (perturbations about the previous) \cite{HallHaw} also closes, while providing a fuller modelling of classical and quantum Cosmology.  

\m 

\n 3) Full GR itself closes \cite{DiracAlg1, DiracAlg2, T73}, in the form of the {\it Dirac Algebroid}, schematically 
\be
\mbox{\bf \{} ( \scM_a    \, | \,    \d L^a   ) \mbox{\bf ,} \, ( \scM_b    \, | \,    \d M^b    ) \mbox{\bf \}}   \es  
              ( \scM_a    \, | \, \, [ \d L, \d M ]^a )                                                                      \m ,
\label{Mom,Mom}
\ee
\be
\mbox{\bf \{} (    \scH    \, | \,    \d J    ) \mbox{\bf ,} \, (    \scM_a  \, | \,    \d L^a    ) \mbox{\bf \}}  \es  
              (    \pounds_{\d\underline{L}} \scH    \, | \,    \d J    )                                                    \m , 
\label{Ham,Mom}
\ee
\be 
\mbox{\bf \{} (    \scH    \, | \,    \d J    ) \mbox{\bf ,} \,(    \scH    \, | \,    \d K    )\mbox{\bf \}}      \es  
              (    \scM_a \mh^{ab}   \, | \,    \d J \, \overleftrightarrow{\pa}_b \d K    )                                 \m , 
\label{Ham,Ham}
\ee  
for $( \m | \m )$ the integral-over-space functional inner product, 
differential-geometric commutator Lie bracket $[ \m , \m ]$, 
and TRi-smearing functions $\d J, \d K, \d \u{L}$ and $\d \u{M}$.   

\m 

\n The third equation's right-hand-side containing the inverse of the spatial 3-metric $\mh^{ab}$ -- a function -- 
is why this is an {\sl algebroid} \cite{Bojowald} rather than an algebra.
Because of this, we introduce the portmanteau {\sl Lie algebraic structure}, to jointly encompass Lie algebras and Lie algebroids. 

\m 

\n Its containing $\scM_a$ as well establishes $\scM_a$ to be an {\sl integrability} of $\scH$ \cite{MT72}.
By this, neither GR's $\scH$, nor its underlying Temporal Relationalism, can be entertained without $\scM_a$ or its underlying Configurational Relationalism.   
Relational Mechanics is moreover even less coupled (Sec 2.6), by which Temporal and Configurational Relationalism can be entertained piecemeal there.

\subsection{GR Observables}
%
{            \begin{figure}[!ht]
\centering
\includegraphics[width=0.27\textwidth]{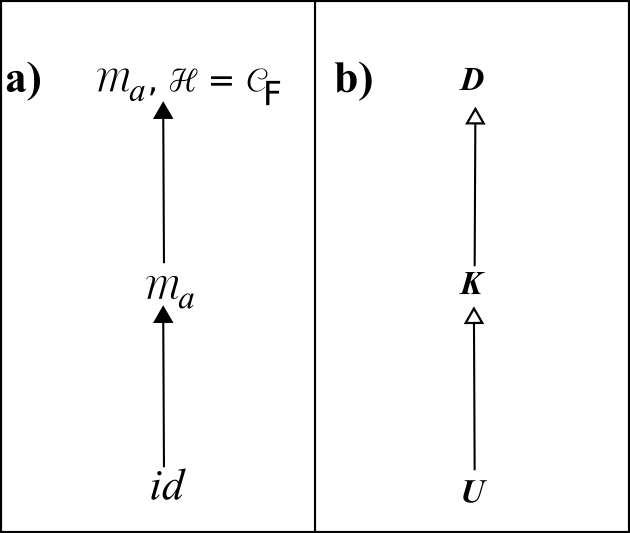}
\caption[Text der im Bilderverzeichnis auftaucht]{ \footnotesize{Chains of GR algebraic structures of a) constraints and b) observables.} }
\label{GR-Obs}\end{figure}            }

\n GR's constraint subalgebraic structures support 3 notions of observables as per Fig \label{G-Obs}.  

\m 

\n 1) Unrestricted observables $\sbiU(\bh, \bp)$. 

\m 

\n 2)  Dirac observables $\sbiD(\mbox{True}, \Pi^{\sT\sr\su\se})$, for true degrees of freedom True: functions commuting with both $\scM_a$ and $\scH$. 

\m 

\n 3)  Kucha\v{r} observables $\sbiK(\bG, \Pi^{\sG})$ are supported for GR \cite{K93} as functions commuting woth $\scM_a$, since $\scM_a$ closes by (\ref{Mom,Mom}) \cite{MT72}. 
This notion of observables does not however have theory-independent significance. 
This role is, rather, replaced by whatever non-extremal elements the bded lattice of notions of observables of a theory happens to possess.

\subsection{GR Rigidity, including Spacetime Construction}

\n This an independent development to Expression in terms of Observables, as indicated by the fork in Fig 1.b). 
Though if new consistent theories emerge in this way, one then needs to plug each of these into the observables-determining procedure. 

\m 

\n While initially conducted using the Dirac Algorithm, consistently resolving the Problem of Time requires use of the TRi Dirac-type Algorithm.  
Moreover, there is now some evidence that Lie's machinery is already capable of finding rigidities.
We thus have a new notion of Tri-Lie algorithm whose domain is any brackets algebra.
So both sides of the fork at Constraint Closure remain within Lie's world (previously called Dirac's world).  
(If Cartan's solutions need be appended where distinct to Lie's, then Lie's dominance over the middle of the Problem of Time would be down by half a facet).
The remaining fork starts by taking {\sl a family of} candidate theories through the TRi Dirac-type Algorithm. 

\m 

\n In this regard, let us consider the following ansatz \cite{AM13} for a family of candidate geometrodynamical theories.  
\be
{\cal S}^{w,y,a,b}  \es  \iint \d^3x \sqrt{\mh} \sqrt{a \, {\cal R} + b} \, \d \ms_{w,y}                \m ,
\label{trial} 
\ee
where $\d \ms_{w,y}$ is built out of the usual $\d - \pounds_{\d\u{\sF}}$ and the more general if still ultralocal (no $\pa_a\mh_{bc}$ dependence) supermetric 
\be 
\mM^{abcd}_{w,y} \:=   \{\mh^{ac} \mh^{bd} - w \, \mh^{ab} \mh^{cd}\}/y                      \m .
\ee  
Its densitized inverse's components are 
\be
\mN_{abcd}^{x,y} \:= y\{  \mh_{ac} \mh_{bd} -  x \,\mh_{ab} \mh_{cd}/2 \}/\sqrt{\mh}  \m , 
\ee 
for  
\be 
x \:= \frac{2 \, w}{3 \, w - 1}                                                                                                                     \m .   
\ee
The parametrization by $x$ is chosen for GR to be the $w = x = y = 1$ case. 
%
%
The quadratic primary constraint is now 
\be
\scH_{x,y,a,b}  \:=  \mN_{abcd}^{x,y} \mp^{ab} \mp^{cd} - \sqrt{\mh} \{ a \, {\cal R} + b \}  
                \es  0                                                                                                                              \m .  
\label{H-trial}
\ee
The self-Poisson bracket of this is  
%
%
\be 
\mbox{\bf \{}           (    \scH_{x,y,a,b}    \, | \,    \d \mJ    ) \mbox{\bf , } \, (    \scH_{x,y,a,b}    \, | \,    \d \mK    ) \mbox{\bf \}}  
\es  a \, y \,          (    \scM_i                \, | \,    \d \mJ \, \overleftrightarrow{\pa}^i \d \mK    ) 
\m + \m  2 \, a \, y \{1 - x\} (    \mD_i \m p \, | \, {\d \mJ} \, \overleftrightarrow{\pa}^i {\d \mK}   )                                               \m .
\label{Htrial-Htrial}
\ee
Relative to (\ref{Ham,Ham}), this picks up an {\it obstruction term to having a brackets algebraic structure}. 
This has four factors, each providing a different way in which to attain consistency. 
GR follows from the third of these setting $x = 1$ \cite{BFO} (see e.g.\ \cite{AM13, ABook} for the other three cases).    
These are very few consistent options as compared to the input family of candidate theories, by which GR is established to be a {\sl rigid} theory.  

\m 

\n When working with minimally-coupled matter terms as well, the above obstruction term is accompanied by a second obstruction term \cite{BFO, AM13, ABook}.
Its factors give the ambiguity Einstein faced of whether the universal relativity is locally Lorentzian or Galilean (Carrollian featuring as a third option). 
This is now moreover realized in the {\sl explicit mathematical form} of a string of numerical factors 
in what would elsewise be an {\sl obstruction term to having a brackets algebraic structure of constraints}.  
In particular, the GR spacetime solution to the first obstruction term is now accompanied by the condition 
from the second obstruction term that {\sl the locally-Lorentzian relativity of SR is obligatory}.  
This can be viewed as all minimally-coupled matter sharing the same null cone because each matter field is separately obliged to share {\sl Gravity}'s null cone.

\subsection{Spacetime's own Relationalism}

\n Having constructed GR spacetime $\bFrM$, one needs to take into account that it has a relationalism of its own: Spacetime Relationalism.
This is with respect to spacetime automorphism group 
\be 
\lFrg_{\sS} = Diff(\bFrM) \m :
\ee 
unsplit spacetime 4-diffeomorphisms. 

\m 

\n This has its own version of closure -- now not Constraint Closure but Generator Closure of $Diff(4)$'s generators, 
\be
\mbox{\bf |[} (    {\cal D}_{\mu}    \, | \,    X^{\mu}) \mbox{\bf ,} \, (    {\cal D}_{\nu}|Y^{\nu}    ) \mbox{\bf ]|} 
\es  (    {\cal D}_{\gamma}    \, | \, \, [X, Y]^{\gamma}    ) \m .  
\label{Lie-2}
\ee
for smearing variables $X^{\mu}$ and $Y^{\mu}$.
\n And of observables: the {\it spacetime observables} $S$ functions on spacetime, which are $Diff(\Frm)$-invariant,i.e.\ Lie-brackets-commute with $Diff(\bFrM)$'s generators, 
\be
\mbox{\bf |[} (    {\cal D}_{\mu} \, | \,  \mY^{\mu})  \mbox{\bf ,} \, (    \iS_{\sfQ} \, | \, \mZ^{\sfQ}    ) \mbox{\bf ]|}  \es 0 \m \label{Sp-Obs} 
\ee
for smearing variables $\mY^{\mu}$ and $\mZ^{\sfQ}$.

\subsection{Refoliation Invariance}
%
{            \begin{figure}[!ht]
\centering
\includegraphics[width=0.7\textwidth]{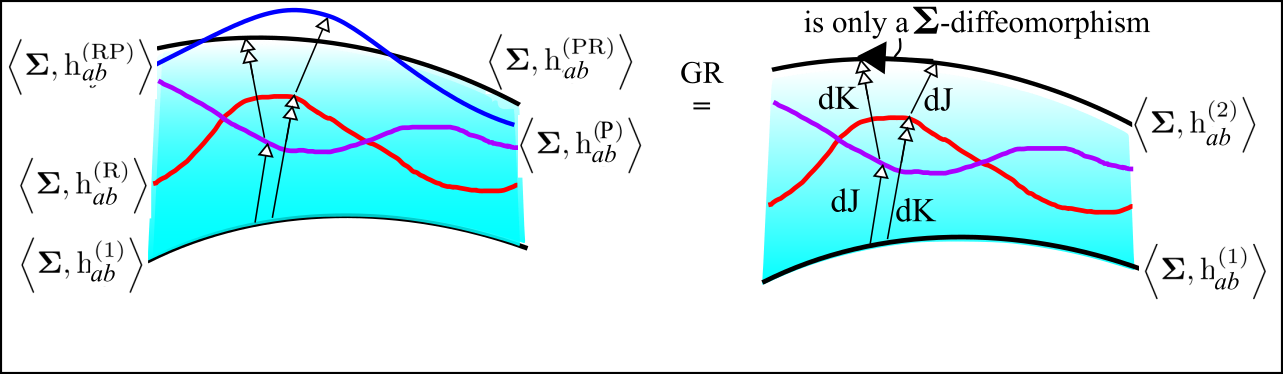}
\caption[Text der im Bilderverzeichnis auftaucht]{ \footnotesize{a) Posing Refoliation Invariance: is going from spatial hypersurfaces (1) to (2) 
via the red (R) intermediary hypersurface physically the same as going via the purple (P) intermediary surface? 
If so, the blue and black hypersurfaces would coincide. 
b) But for GR, (\ref{Ham,Ham}) gives that the black and blue hypersurfaces can at most differ by a spatial diffeomorphism, and so must coincide.} }
\label{Refol-Inv}\end{figure}            }

\n Let us finally address Foliation Independence.  
We formulate the corresponding foliation kinematics along the lines of Isham \cite{I93}. 
Though again a new TRi version (rather than the original ADM-like version) of this kinematics 
is required to retain compatibility with Temporal Relationalism: {\it TRiFol} \cite{TRiFol} (`Fol' standing for foliations). 

\m 

\n This scheme moreover has Refoliation Invariance guaranteed, by a TRi version of {\it Teitelboim's approach} \cite{T73}: 
noting that within the Dirac algebroid of GR (\ref{Ham,Ham}) is none other than a local algebraic formulation of Refoliation Invariance (Fig \ref{Refol-Inv}). 
Conversely, resolving Foliation Independence by Refoliation Invariance {\sl requires} an algebroid to keep track of all the foliations.  
While not originally formulated in a TRi manner, this can be arranged \cite{TRiFol}, via the emergent-instant counterpart of Fig \ref{Strutty}.b)'s kinematics.  

\m 

\n This also lies within Lie's arena, though various subsequent developments are required: 
Dirac, Teitelboim, late-70's further formulations in part by Kucha\v{r} and reviewed within \cite{I93}.

\section{Quantum Version of Problem of Time and its local resolution}

\n I proceed via giving a {\it TRiCQT} \cite{ABook}: a TRi variant of Isham's Canonical Quantum Theory \cite{I84}.  

\m

\n Kinematical quantization \cite{I84} comes first, so as to provide us with incipient operators. 
This is an analogue of finding unrestricted observables, which we now denote by $\widehat{\sbiU}$.
This is now a nontrivial endeavour since there are substantial restrictions on which classical functions of $\biQ$, $\biP$ can be consistently promoted to quantum operators.

\m 

\n Everything else follows in the classical version's order, which thus conceptually and technically pre-empts the quantum version.  

\m 

\n We next promote the classical first-class constraints arising from Configurational Relationaism and Temporal Relationaism to quantum constraints, 
\be 
\bFlin     \m \longrightarrow \m  \widehat{\bFlin}     \m ,
\ee
\be 
\Chronos  \m \longrightarrow \m  \widehat{\Chronos}  \m .
\ee
These act on the {\it wavefunction of the Universe} $\Psi$. 

\m 

\n We subsequently solve 
\be
\hat{\bFlin} \, \Psi = 0
\ee 
(if any). 
We next solve 
\be 
\widehat{\Chronos} \, \Psi = 0  \m , 
\label{Hat-Chron}
\ee
the GR case of which is the notorious {\it Wheeler--DeWitt equation} \cite{Battelle, DeWitt67}
\be 
\widehat{\scH} \, \Psi = 0   \m . 
\ee  
This is one of the places in which the most well-known Problem of Time facet -- the Frozen Formalism Problem -- appears, 
since one would be expecting a time-dependent wave equation at this point (for some notion of time).
Solving (\ref{Hat-Chron}) is termed {\it dynamical quantization} \cite{I84}.

\m 

\n Most of the subsequent work offered in \cite{ABook} is for Semiclassical Quantum Gravity, for which 
\be 
\Psi  \es  \me^{i \, {\cal S}(h)} \, | \, \chi(h, \, l) \, \rangle   \m ,
\ee 
for $h$ slow heavy and $l$ fast light degrees of freedom.
This suffices from the moderate standpoint that this is at most what shall be observed in coming decades. 
The fairly well-known notion of emergent semiclassical time $t^{\sss\se\sm}$ \cite{I93} additionally admits a Machian interpretation \cite{ABook}. 
This needs to be `found afresh' at the quantum level, rather than continuing to use the classical $t^{\se\sm}$. 
This is for the furtherly Machian reason that abstracting GLET from STLRC is sensitive to partial or total replacement of classical change by {\sl quantum change}, e.g.\ of 
\be 
\d l \m \mbox{ by } \m \d|\chi(h, \, l)\rangle \m .   
\ee 
\n We next need to check the hatted version closes 
\be 
\mbox{\bf [} \widehat{\scC}_F \mbox{\bf ,} \, \widehat{\scC}_{F^{\prime}} \mbox{\bf ]} \, \Psi = 0 \m .
\ee 
[Isham and Kucha\v{r} termed the corresponding facet {\it Functional Evolution Problem}, with particular reference to its field-theoretic incarnation.] 
The Configurational to Temporal Relational split version of this is 
\be 
\mbox{\bf [} \widehat{\Flin}_L \mbox{\bf ,} \, \widehat{\Flin}_{L^{\prime}} \mbox{\bf ]} \, \Psi = 0           \m ,  
\ee
\be 
\mbox{\bf [} \widehat{\bFlin} \mbox{\bf ,} \,  \widehat{\Chronos} \mbox{\bf ]} \, \Psi = 0       \m , 
\ee
\be 
\mbox{\bf [} \widehat{\Chronos} \mbox{\bf ,} \, \widehat{\Chronos} \mbox{\bf ]} \, \, \Psi = 0  \m .  
\ee
\n We need to upgrade from $\hat{\sbiU}$ to $\hat{\sbiD}$ observables: quantities commuting with all quantum constraints $\widehat{\sbcC}$
This amounts to `finding afresh' quantum observables, 
This is as opposed to trying to promote classical observables, which would face three obstacles. 
Firstly, the change from Poisson brackets to commutators. 
Secondly, the restrictions already imposed by kinematical quantization. 
Thirdly, the promotion of classical constraints $\sbcC$ to their quantized forms $\widehat{\sbcC}$.  

\m 

\n One can next pose semiclassical rigidity, and so on.

\section{Global version of the Problem of Time}\label{Strata}

Each facet resolution given above is local. 
It would be preferable if all of the Background Independence aspects, 
                                     resultant Problem of Time facets, and 
									 strategies to resolve these,           were treated in a globally well-defined manner. 
We are however far from this goal, which involves many -- and often distinct -- uses of the word `global', some of which have so far not even been explained in the literature.   

\m

\n This can be local in               space, 
                        configuration space, 
                                phase space, 
								spacetime, 
						space of spacetimes, 		
						or constraints, generators or observables algebraic structures over one of the preceding...
%
						
\m 

\n The concept of locality in question is in the topological sense of neighbourhoods.

\subsection{Configurational Relationalism is globally underscored by how $\lFrg$ is found}\label{Cl-Glob-G}

\n As we already emphasized, infinitesimal group generators are formulated in terms of the Lie derivative, 
as for instance in (\ref{Lie-Drag-1}) or (\ref{Lie-Drag-2})'s correction terms.  

\m 

\n Prima facie, then, systematically solving (generalized) Killing equations for these infinitesimal group generators raises no eyebrows, 
as these equations are also in terms of Lie derivatives.
And yet these (generalized) Killing equations are sensitive to the underlying carrier space $\FrC^d$'s topology. 
There are two reasons for this. 
The first is that Lie derivatives are capable of being either locally or globally defined. 
This is because they are defined with respect to a vector field, and one may then distinguish between this being defined in a neighbourhood of the origin or over the whole manifold.
The second is that (generalized) Killing equations are elliptic PDEs.  
That (generalized) Killing equations do indeed `feel' the underlying space's topology is already clear from 
their having rather different solutions on flat $\mathbb{R}^d$ and flat $\mathbb{T}^d$ (c.f.\ Article I).

\m 

\n Because of this, dealing with Configurational Relationalism has a preliminary global portion. 
Namely,  systematically solving the corresponding generalized Killing equation for a basis of generators for the automorphism group's Lie algebra.
Only then does one enter local workings: infinitesimally correcting velocities in the geometrical Principles of Dynamics action with these automorphism algebra generators, 
and then following through as per Sec 2.3, or similarly with changes as per Sec 2.5.

\subsection{Configurational Relationalism gives Stratified Manifolds}\label{CR-StratMan}

At the level of configuration spaces, for $N$-point-or-particle models, implementing Configurational Relationalism in general produces a stratified manifold (Article II).  
Some are LCHS and some are merely Kolmogorov.

\m 

\n Returning to one of Article II's main examples, in 3-$d$ Euclidean Relational Mechanics, 
elimination of the rotational auxiliary breaks down for collinear configurations since these have singular inverse inertia tensors.
The collinear configurations moreover constitute a separate stratum.  

\m

\n In the GR counterpart, the $\bh$ with nonzero Killing vectors -- spatial metrics with symmetries -- form nontrivial stratification.  
DeWitt gave a conceptual account of this in \cite{DeWitt70}, with Fischer concurrently providing a superbly detailed technical account \cite{Fischer70}.  

\m 

\n This is the arena of Fischer's strata-unfolding proposal \cite{Fischer86} mentioned in Article II.

\m

\n In GR's Thin Sandwich case of Best Matching, localization away from both 1) potential factor zeros and 2) metrics with nontrivial Killing vectors recur.
In particular, both appear in Bartnik and Fodor's Thin Sandwich Theorem \cite{BF} as locality restrictions.
2) can moreover be interpreted as an instance of strata excision (c.f.\ Article II).  

\m 

\n One distinction between Shape Theory's and GR's strata is that while Shape Theory's realize Article II's         frontier condition, 
                                                                        GR's           realize instead the `inverse frontier condition'  \cite{Fischer70}.

\m 

\n Another is that the full GR's configuration spaces are infinite-dimensional.  
While some work on infinite-dimensional stratifolds has already been contemplated \cite{Ewald, Tene, KT15}, 
the function spaces in use are not yet of the required kind to tackle GR $\Superspace(\bupSigma)$, namely \cite{Giu09, ABook} Fr\'{e}chet spaces.   
%

\m 

\n Zeros and strata place local limitations on at least the current-era straightforward solving procedures for `best matched' values of the $\lFrg$-auxiliary variables.
Both of these moreover reflect features that do not enter the generalized Killing equation itself. 
Namely, what choice of potential factor is made over the original configuration space, and the structure of the r-quotient configuration space.

\m 

\n The strata in particular retain information about the ab inito carrier space's symmetry and topology. 
This occurs via strata picking up nontrivial isotropy groups alias stabilizers from quotienting out the automorphism group. 
This (partial) `memory' of the carrier space's symmetry group is moreover also a (partial)`memory' 
of the carrier space topology since the automorphism group is obtained by solving the (generalized) Killing equation, which `feels' this as per Sec 5.1.  

\m

\n In retrospective, this failure in general to erase all topological and symmetric memory of the ab initio carrier space is unsurprising, 
through the identification of `Best Matching' as being implemented via local Differential Geometry. 
Such methodology comes with no guarantees of being able to banish topological features of the ab initio carrier space. 
Nor all symmetric features, since these themselves have a partly-global character as per Sec 5.1.
For now, we offer two different directions toward erasing this relationally undesirable ab initio absolute space's symmetry and topology information in Secs 5.4 and 5.5,  
while keeping two more in reserve for future occasions.

\subsection{Global Temporal Relationalism}\label{Cl-Glob-tem}

Representing motions by geometrical actions whose solutions are geodesics on configuration space $\FrQ$ is prima facie attractive.
This works out locally if one considers the geometrical action.  
Such actions produce $\Chronos$ constraints and emergent Machian times $t^{\se\sm}$.

\m 

\n Attempting a global version, however, meets three impasses.  

\m 

\n Impasse 1) Zeros of the potential function, $W = E - V$ for Mechanics or $2 \, \slLambda - {\cal R}$ for GR, at least prima facie obstruct progression of dynamical trajectories. 
This is additionally qualitatively different in each of these two cases due to metric definiteness versus metric indefiniteness; see \cite{Epi-B} for an outline.  
This moreover causes the formula for $t^{\se\sm}$ to contain a division by zero.

\m 

\n Impasse 2) the prima facie attractiveness of conceiving of physics in terms of geodesics on configuration also merits a geometrical challenge. 
For the underlying chain of thought is that configuration spaces are manifolds, and geodesics thereupon are a straightforward matter.  
But the reality is that configuration spaces are in general {\sl stratified manifolds}, and geodesics thereupon are not a straightforward matter. 

\m 

\n i) Geodesics need to be computed stratum-by-stratum, since affine and metric structure are only assigned in this manner.

\m 

\n ii) There is the further matter of how to continue geodesics which strike a boundary between strata.  

\m 

\n Impasse 3) The GLET modelling assumption is itself local. 
For instance, its STLRC could be based on a heavy--light split of degrees of freedom, but what is heavy and what is light could be local over space and/or configuration space.  

\m 

\n Temporal Relationalism's locality is thus an open set within a single stratum, containing no potential space zeros and throughout which a single GLET specification holds. 

\m  

\n Globalizing this requires a stratum-traversing procedure, 
                             a zero traversing procedure (of which \cite{SS94} provides a simple example), and 
							 a patching procedure between adjacent patches' GLETs.

\subsection{Triviality of Kendall-Type Relational Theories on generic carrier space}

\n Article I and II's arguments allow for this second option as well. 
Both horns of this dilemma are moreover difficult in different ways, in the manner of Scylla versus Charybdis, referring to having no choice but 
to sail through the abode of one mythological monster or the other. 
Our dilemma moreover carries considerable Absolute versus Relational Debate significance.

\m 

\n{\bf Scylla} Consider ab initio absolute-space symmetry. 
This in general leads to complicated stratified manifolds, which in turn exceed fibre bundles' capabilities, necesitating rather general bundles, presheaves or sheaves thereover.  
This occurs because the model's strata remember some of the symmetry information about the ab initio absolute space, 
even after the corresponding automorphism group has been quotiented out \cite{A-Generic}. 
These stratified manifolds can moreover be merely-Kolmogorov separated rather than Hausdorff-separated.

\m 

\n{\bf Charybdis} Consider instead a generic asymmetric ab initio absolute space, e.g. through incorporating small defects to nullify any symmetry that would elsewise be present. 
This case has no strata or need for more than fibre bundles as induced from stratification. 
However, it also lacks computational tools since so many of these are rooted in symmetry 
(e.g.\ \cite{ABook} conservation, perturbation expansions, special functions, special forms taken by metrics...)

\m

\n Models are moreover also capable of remembering ab initio absolute space's topology even after the corresponding automorphism group has been quotiented out \cite{ACirc}.

\m 

\n These two kinds of memory are perhaps not so surprising. 
Quotienting out the automorphism group is a {\it differential-geometric} matter, as is clear from Best Matching starting with Lie derivative corrections.  
This comes with no guarantees about controlling the model's topology.  
The differential-geometrical category is moreover not in general well-behaved under quotients. 
By this, symmetry information can remain topologically recorded as strata even as quotienting out the automorphism group erases differential-geometric symmetry information. 
In this way, the {\sl the relationalists' quest to free Physics of any imprint of absolute space} with merely differential-geometric tools such as Best Matching 
succumbs to the Scylla of `stratification from symmetry'.
Further attempts to continue this route are forced to work at the topological level.  
 
\m 

\n A dilemma of this kind also occurs in the GR setting. 
In fact, in GR this was already pointed out two decades ago by Fischer and Moncrief \cite{FM96}, 
to the Shape Theory version being new to this year (\cite{A-Generic} and the current series). 

\m 

\n The generic 3-manifold $\bupSigma$ has no Killing vectors, so the corresponding $\Superspace(\bupSigma)$ is unstratified and thus  
\be
\Superspace(\bupSigma) \m \mbox{ is just a manifold}   \m .  
\ee
One is however unable to proceed with almost any standard GR calculations, 
out of almost all of these being rooted in \cite{MacCallum} symmetric rather than generic $\bupSigma$ (or expansions thereabout).

\subsection{Sum over all carrier spaces' topologies}

A way of erasing all memory of topology and symmetry of the ab initio absolute space is to `sum over all possible topologies' or `extremize over all possible topologies'. 
We note this to be the `all' completion of the topological manifold level equivalent of the $\lFrg$-act, $\lFrg$-all method.
A priori, the `sum' example in particular is in practice very (and may in principle be arbitrarily) computationally expensive. 
But especially the `extremize' example lends itself to selection principles and approximation methods. 

\m 

\n This approach on the one hand includes the generalization of GR to allow for spatial topology change \cite{GH92}, 
           while on the other hand suggesting the new arenas of `sum over all absolute spaces' Relational Mechanics and `sum over all carrier spaces' Shape Theory.

\subsection{Constraint Closure and Expression in Terms of Observables}\label{Cl-Glob-CC} 

\n Constraint algebraic structures form a copresheaf (with the extension maps of Fig 3.a).

\m 

\n Expression in Terms of Observables is the dual presheaf \cite{ABook, DO-1} (with the restriction maps of Fig 3.b).

\m 

\n Another global issue is that observables are often presented as coordinate functions, 
which are not defined globally on curved manifolds such as $\FrQ$ or $\Phase$ but rather just in coordinate patches. 

\m 

\n A further global issue is that notions of types of observables themselves are themselves in general only construed to hold locally.
This is because they are defined by brackets, which -- paralleling the previous Section's Problem 2) -- in general only hold in a local-in-time-and-space slab.  
Moreover, writing out the defining brackets now explicitly gives functional DEs for the observables, which in general only carry local guarantees for solutions.  
All in all, patching observables is at the level of functional DE solutions rather than just of the plain configuration space's Differential Geometry.  

\m 

\n Let us term observables that are local in time and space, respectively {\it fashionables}: a term introduced by Bojowald et al \cite{Bojo1, Bojo2, Bojo3}. 
These are fitting nomenclature for local versions of these concepts: `fashionable in Italy', `fashionable in the 1960s'.  

\m 

\n There are moreover patching procedures available for dealing with the above global issues.

\subsection{Global Rigidity, Spacetime Reconstruction and Spacetime Relationalism}

\n At the global level, rigidity results from brackets algebras have cohomology and obstruction underpinnings.  

\m 

\n The GR Cauchy evolution input into Spacetime Construction is restricted moreover by the current insufficiently developed state of Global Analysis.

\section{Justifying local resolutions in the first place}

Article II pointed to a subcase of Relational Theories having LCHS locally compact Hausdorff second-countable) relational spaces.
The current Article concerns having such spaces, or some suitable weaker condition, as a {\bf Selection Principle} among models with ab-initio symmetric absolute spaces. 
This rests furthermore on Mumfordian excision of unstable strata not being justifiable in the physical setting, as per Article II.  

\m 

\n  With 
\be 
\mbox{LCHS} \m  \Rightarrow \m \mbox{P} \m \mbox{ (paracompactness)}    \m .    
\label{LCHS->P}
\ee 
as well, our concrete Selection Principle is as follows. 
\be 
\mbox{`r-spaces are to be HP (Hausdorff paracompact) spaces'}  \m . 
\ee 
This rests on {\sl adopting a local approach having a particularly strong justification for HP spaces}, for the following reasons.  

\m 

\n 1) HP spaces admit {\sl partitions of unity} \cite{Munkres}. 

\m 

\n 2) It follows that {\sl bump functions} are well adapted to this setting \cite{Munkres, BG88}. 

\m 

\n 3) Partitions of unity moreover prop up the familiar kind of theory of integration \cite{AMP, Lee2}. 

\m 

\n 4) HP spaces also support differential structures, e.g.\ developed on Article II's Differential spaces in \cite{SniBook}. 
Differential Geometry is moreover aided by bump functions.

\m 

\n 5) It follows from 2-4) that PDEs \cite{H90} and variational principles can be posed on HP spaces. 
PDE Theory is also aided by bump functions.  

\m 

\n 6) These in turn are what conventional continuum-like Theories of Physics require for substantial and conventional developability. 

\m

\n 7) We shrink the local region if necessary; the Shrinking Lemma \cite{Munkres, Wedhorn} is available for HP spaces, by which multiple local criteria can be compatibilized.

\m 

\n 8) We still have to deal with associated spaces over HP stratified manifold r-configuration spaces $\w{\FrQ}$, 
as some of our locality conditions are posed in these rather than in $\w{\FrQ}$.
(Pre)heaves are a universal language for such a notion of `over'.  
For HP spaces, moreover, sheaf cohomology also reduces to \cite{Wedhorn} the simpler, older and more widely studied \v{C}ech cohomology.
This is of further relevance since some of the local criteria are on associated spaces, such as observables algebras over reduced phase space or PDE solution spaces. 
For these, placing (pre)sheaves over our reduced spaces is a mathematical means of incorporating and jointly treating these local criteria. 

\m 

\n Moreover 8) {\sl itself} follows from \cite{Wedhorn} HP spaces admitting 1)'s partitions of unity and 7)'s Shrinking Lemma. 
For ease of finding further discussion and results in this area, 
we note that the technical name for sheaves that are simpler in this way is {\it soft sheaves} \cite{Wedhorn, Wells}.  

\m 

\n All in all, 7) and 8) strongly justify being able to isolate local treatment as a meaningful subproblem of an ab initio global problem. 
In the present case, this refers to the local Problem of Time being a consistent subproblem of the full (global) Problem of Time, whenever the underlying spaces are HP. 
Which, in our particular local resolution of the Problem of Time, requires HP reduced configuration spaces.
This is since reduction down to such features very early (as per Sec 2.3, 2.5) on in its network of moves for addressing the seven local classical facets of the Problem of Time.    

\m 

\n The corresponding GR reduced space -- Wheeler's superspace -- is moreover also HP. 
Both superspace and Kendall's own approach both proceed via a distinct general Topological Spaces result 
\be 
\mbox{metrizability} \m  \Rightarrow \m \mbox{HP} \m .
\label{M->HP}
\ee 
in place of (\ref{LCHS->P}).
Hausdorffness is clear, by use of balls to `house off', whereas paracompactness is proven in e.g.\ \cite{Munkres}.  
The above justification of localization thus transcends to the Problem of Time's most traditional GR-as-geometrodynamics setting as well.

\m 

\n It remains to outline what 1), 2) and 7) consist of, for those readers less familiar with Differential Geometry or Topology.

\m 

\n Given a cover $\{U_{\alpha}\}_{\alpha \in \sFrA}$ of a topological space $X$ by open sets, a {\it partition of unity} dominated by this cover is family of continuous functions 
\be 
\phi_{\alpha}: X \m \longrightarrow \m [0, 1]
\ee
such that the following properties hold.

\m 

\n i) The support of $\phi_{\alpha}$, 
\be
\mbox{Supp}(\phi_{\alpha}) \subseteq U_{\alpha} \m \mbox{ for each }  \m \alpha \in \FrA \m .
\ee
[For $f$ a function $Supp(f)$ is the set of values $y$ in $f$'s domain such that $f(y) \neq 0$.]

\m 

\n ii) $\{ \mbox{Supp}(\phi_{\alpha}) \}_{\alpha \in \sFrA}$ is locally finite.

\m 

\n iii) We have the unity condition  
\be 
\sum_{\alpha \in \FrN} \phi_{\alpha}(x) = 1 \m \mbox{ for each } \m x \in X
\ee 
for $\FrN$ the $\FrA$ for which $\phi_{\alpha}(x)$ is nonvanishing in a neighbourhood of $x$. 
[This $\FrN$ is finite by ii), by which our sum is indeed well-defined.]

\m 

\n A {\it bump function} is a smooth function that takes the value 0 outside of some region $\FrU$ and 1 on another region $\FrV \subset \FrU$. 

\m 

\n See Fig \ref{Bumps} for concrete examples on $\mathbb{R}$ and $\mathbb{R}^2$.  
%
{            \begin{figure}[!ht]
\centering
\includegraphics[width=1.0\textwidth]{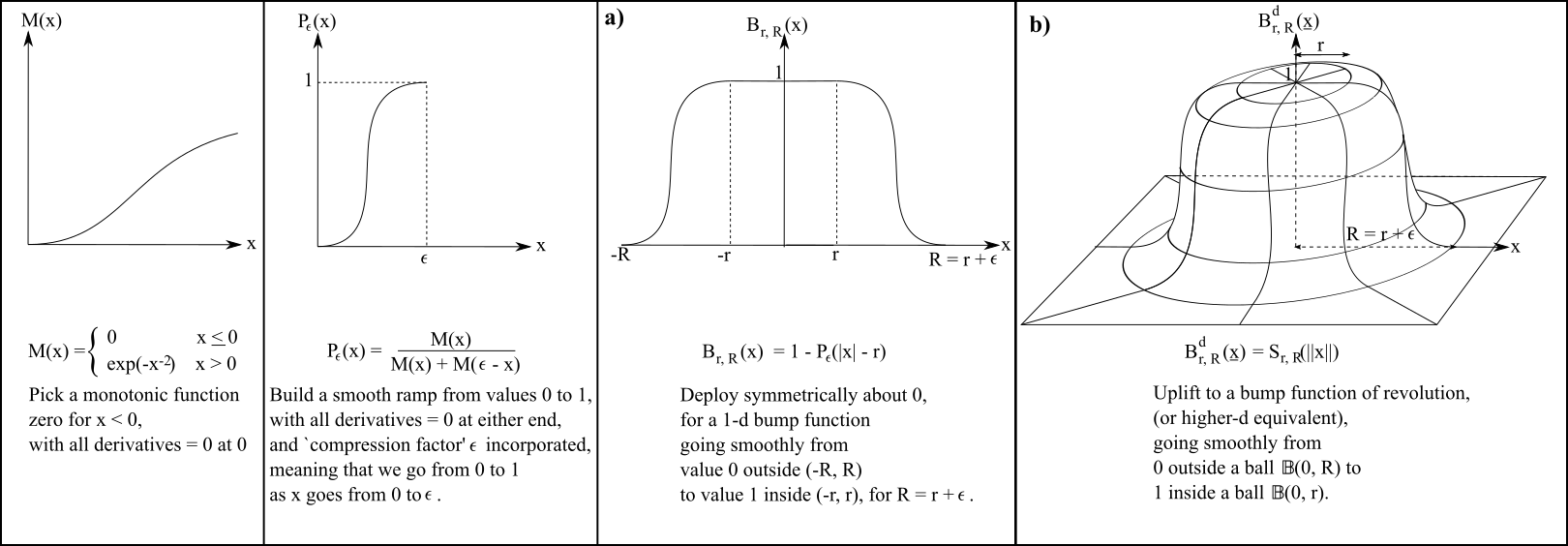}
\caption[Text der im Bilderverzeichnis auftaucht]{ \footnotesize{Bump functions a) on the interval $(-R, R)$ in 1-$d$. 
b) On the open ball $\mathbb{B}^d(0, R)$, using the same function as a radial profile for a surface of revolution bump function.} }
\label{Bumps}\end{figure}            }

\m 

\n The {\it Shrinking Lemma} in question is as follows.  

\m

\n Let $X$ be an HP space, and $\{U_{\alpha}\}_{\alpha \in \sFrA}$ be a cover of $X$ by open sets.
Then there exists a locally-finite cover $\{V_{\alpha}\}_{\alpha \in \sFrA}$ such that for each $\alpha$ the closure $\overline{V}_{\alpha} \subset U_{\alpha}$.

\m 

\n For those readers whose specializations do make considerable use of topological spaces, moreover, we note that this is not the only Shrinking Lemma.  
The further topological properties of {\sl normality} \c{Munkres} and {\it metacompactness} \cite{Nagata} are relevant here, 
and might conceivably support weakenings of, or alternatives to, the current Article's HP-space Selection Principle. 
Note moreover that 1) to 8)'s needs are collectively more stringent than the question of which topological spaces admit a Shrinking Lemma...

\section{Comparative Background Independence at the global level}\label{Cl-Top}

\subsection{Local treatment of flat geometrical levels of structure}\label{Flat}

The Author has shown that aspects 1) to 4) of our local resolution carry through \cite{AMech, AObs2, ABook} for flat-space Affine, Equireal and Conformal Relational Mechanics. 
In the projective case, subtleties similar to \cite{MP05}'s in Shape Statistics have held up the appearance of a Projective Relational Mechanics.
Aspects 1) to 4) also hold for the isometric theory on the flat torus $\mathbb{T}^d$ \cite{ATorus}, with the distinction that this case's resolution is simpler that all the other cases considered.  
These examples all rest on flat space, pace projectivization and topological identification in the last 2 cases. 
Alongside Spherical Isometric Relational Mechanics \cite{ASphe}, these examples are moreover $N$-point-particle theories.  

\m 

\n At level of pure geometry, have a generator brackets analogue of the Dirac Algorithm.
For the general homogeneous-quadratic generator ansatz \cite{A-Brackets}
\be 
Q^{\st\sr\si\sa\sll}_{\mu, \nu \, A}  \:=  \mu ||x||^2\pa_A + \nu \, x_A (\u{x} \cdot \u{\nabla} )                             \m ,  
\ee 
 \be 
\mbox{\bf [} Q^{\st\sr\si\sa\sll}_{\mu, \nu \, A} \mbox{\bf ,} \, Q^{\st\sr\si\sa\sll}_{\mu, \nu \, B} \mbox{\bf ]}  \es  2 \, \mu \, (2 \mu + \nu) ||x||^2x_{[A}\pa_{B]}  \m . 
\ee 
The first strongly vanishing factor corresponds to Projective Geometry and the second to Conformal Geometry: 
$\mathbb{R}^{d \,  \, \geq \, \,  3}$'s two possible choices of `group-theoretically top geometries.  
This result moreover establishes that rigidity is not some province of phase space--Poisson-brackets--Dirac-Algorithm complex of techniques.

\subsection{Local treatment of Differential Geometrical levels of structure}\label{Diff}

\n Differential-geometric counterparts include GR-as-geometrodynamics, as well as weakening to metrodynamical theory and strengthening to conformogeometrodynamical theory.  

\m 

\n Metrodynamics \cite{AM13} has ab initio trivial Configurational Relationalism.
However, $\scH$ from its Temporal Relationalism enforces $Diff(\bupSigma)$-Configurational Relationalism at the level of Constraint Closure, 
so one arrives at GR again, albeit now resting on less initial assumptions. 

\m 

\n Conformogeometrodynamics \cite{York, ABFO, AM13} has maximal
\be 
\mp = 0   \m ,
\ee 
or, more generally, constant-mean-curvature condition, 
\be 
\frac{\mp}{\sqrt{\mh}} = const \m , 
\ee 
This decouples the constraints, permitting solution of $\scM_a$ prior to $\scH$.  
The usual way of solving $\scH$ moreover breaks the spatial conformal inv, returning GR-as-Geometrodynamics, albeit also now having addressed its initial-value problem.  
One can also get confgdyn by choosing the fourth factor in GR's rigidity equation: 
\be 
\mD_a \mp = 0 \m \Rightarrow \m \frac{\mp}{\sqrt{\mh}} = const \m . 
\ee   
Trying to give such an independent theory status, however, runs into trouble with Refoliation Invariance, 
which this Author adheres to, thus discarding such alternative theories as failing to embody the seventh local aspect of Background Independence.

\subsection{Selection principle's global-level restriction for flat space levels of structure}\label{Sel}

\n Globally, however, our Selection Principle of r-spaces being HP selects, as per Article II (in particular its Figure), 
just a narrow band of Relational Theories as technically palatable. 
The excluded theories are moreover not only blocked from global development, but are globally such that considering localization therein is reductio ad absurdum.  

\m 

\n Euclidean and Similarity Relational Mechanics on $\mathbb{R}^d$ survive, as do Isometric Relational Mechanics on $\mathbb{T}^d$, $\mathbb{S}^d$ and $\mathbb{RP}^d$.  

\m 

\n On the other hand, flat-space Affine and Projective Relational Mechanics are out on this count.
It is rather probable that Conformal Relational Mechanics, and Minkowskian Relational Mechanics to all significant levels of automorphism group structure, 
will follow these out as well. 

\m 

\n This is the first way in which our Selection Principle shapes the Comparative Theory of Background Independence.

\subsection{Global Comparative Background Independence among Differential Geometric level theories}

Fischer and Moncrief showed \cite{FM96} that $\CS(\bupSigma)$ has rather tight parallels with $\Superspace(\bupSigma)$'s own global character. 
Namely, manifolds possessing conformal Killing vectors possess separate strata, 
whereas for the generic 3-manifold $\bupSigma$ has no conformal Killing vectors, so the corresponding $\CS(\bupSigma)$ is unstratified and thus  
\be
\CS(\bupSigma) \m \mbox{ is just a manifold}           \m . 
\ee
\n In contrast to Sec 3.6's Thin Sandwich, whose zeros can appear locally in space, 
the rival conformal approach to GR's initial value problem equations are better behaved globally in space.

\m 

\n Our Selection Principle clearly favours space over spacetime approaches, since $Superspace(\bupSigma)$ is HP whereas 

\n $Superspacetime(\FrM)$ is not Hausdorff.  
This is the second way in which our Selection Principle shapes the Comparative Theory of Background Independence.  

\m 

\n Our Selection Principle is not to date known exclude any Differential Geometric level theories. 
This may in fact reflect that humankind is lagging behind with the dynamical structure of affine and projective counterparts to GR.

\subsection{Deeper levels of structure}\label{Deeper}

In the standard Equipped Sets Foundational System of Mathematics, differential structure is underlied by each of the following in turn.  

\m 

\n 1) Topological manifold change \cite{Milnor-h, GH92, Witten88}.

\m 

\n 2) Topological space change \cite{I91, Top-Shapes, Forth}.  

\m 

\n 3) Change of collection of subsets \cite{Forth}. 

\m 

\n 4) Change of underlying set \cite{I91, Forth}.  

\m 

\n Article II outlined the first two of these; the next two are structurally simpler.\footnote{This does not happen in some other Foundational System of Mathematics, 
though we leave this for another occasion.
For instance, topological spaces could be taken to rest upon sheaves, the spaces of which are topoi \cite{Bell}.}  

\m 

\n The {\sl posing} of Temporal Relationalism, 
                       Configurational Relationalism, 
					   Constraint Closure, 
					   Expression in Terms of Observables, 
				   and Rigidity can be made at any level of mathematical structure \cite{Epi-C}. 

\m  

\n But solving Configurational Relationalism, 
            Constraint Closure's consistency, 
			solving for observables 
	    and demonstrating rigidity are case by case for each level.  

\m 

\n The current series' consideration (Sec 7.3) lie within Differential Geometry (as does its parallel in Sec 7.4) so Lie's mathematics remains available.  

\m 

\n Not much is known about topological spaces in the role of physical space.  

\m 

\n A) Manifolds, yes.  

\m 

\n B) Metric spaces, yes \cite{I91, ASoS, ABook}.

\m 

\n C) As regards LCHS spaces, Article II has a theorem guaranteeing that reducing such preserves LCHSness.

\m 

\n But not many specifics are known. 
Not even $N$-point models on manifolds with boundary have been done at the reduced and space of spaces level to the Author's best knowledge. 
Nor has superspace for space with boundary.

\m

\n D) Sets becomes considerably easier to handle, due to structural sparsity \cite{I91}.

\m 

\n Note finally that, while modelling space as a topological space $X$ comes with HP limitations, 
one can imagine distinct models of space, such as using discrete graphs for $N$-point rubber shapes \cite{Top-Shapes} will allow for other kinds of `islands of good behaviour'.

\section{Conclusion}

\subsection{The local resolution of the Problem of Time is based on Lie's mathematics}

\n A local resolution of the full, multi-faceted Problem of Time \cite{K92, I93, Battelle, DeWitt67} was recently put forward \cite{A-Lett, ABook}.  
This is unprecedented in handling all seven of the local facets concurrently, 
these facets having moreover a history of interfering with each other so that solving multiple facets concurrently involves much more than the `sum of its parts'.
A local resolution of the Problem of Time moreover amounts to establishing a local theory of Background Independence \cite{APoT3, ABook}.

\m 

\n \cite{ABook}'s presentation of this local resolution viewed it in three pieces 
(the below numerals labelling the Background Independent aspects corresponding to overcoming the Problem of Time facets).

\m 

\n A) 1) Configurational and 2) Temporal Relationalism serve as constraint providers, and are addressed along lines pioneered by Barbour \cite{BB82, B94I}.

\m 

\n B) The ensuing constraints are to be checked for consistency: 3) Constraint Closure. 
This is addressed along the lines of Dirac's Algorithm \cite{Dirac}. 
It marks the beginning of the realm of brackets algebras, which then forks into two mutually-independent parts. 

\m 

\n 4) Expression in Terms of Observables `What forms zero brackets with the constraints' gives the observables \cite{DiracObs}. 

\m 

\n ii) `Running whole families of candidate theories through a Dirac-type algorithm' gives a rigidity: that GR is one of very few cases surviving this consistency check. 
This moreover also provides a means of, firstly, deriving GR-as-Geometrodynamics specific features from within ab initio entire families of candidate geometrodynamical theories.  
Secondly, of constructing GR-like spacetime -- 5) Spacetime Construction -- from merely spatial structure alongside the underlying demand of consistency.

\m 

\n C) Spacetime constructed, 6) this has its own Relationalism, Closure (now of spacetime automorphism group generators, not constraints) and Observables. 
Spacetime can moreover also be foliated and is Refoliation Invariant,  
with moreover its constraint algebraic structure implying Refoliation Invariance \cite{T73} and thus 7) Foliation Independence.  

\m  

\n The current Article is moreover the first to declare that this Provider--Brackets--Spacetime troika is underlied at the classical level by a single body of mathematics: 
that initiated by Lie.  

\m 

\n This mathematical observation, on the one hand, immediately rings true, due to the largely local nature, and, within Differential Geometry, universal character of Lie's work.  
In concordance, both Relational Mechanics and GR are Differential Geometry level theories.
So a local resolution of the Problem of Time in each of those arenas would be expected to involve Lie's largely local Differential Geometry level mathematics.  
{\sl Lie's mathematics, as the cornucopia of local Differential Geometry, succeeds in providing a full complement of mathematics for a local resolution of the Problem of Time}.  

\m 

\n On the other hand, this mathematical observation is lucid. 

\m 

\n For instance, it permits the following insight into our local resolution of the Problem of Time.   
Firstly, 2)'s changes and 1)'s corrections are formulable as \cite{ABook} Lie derivatives. 
Secondly, Lie brackets based 3) checks on the ensuing constraints, 
                             4) definitions of observables, 
						 and 5) checks again, now on entire families of constraints arising from entire families of candidate theories. 
Thirdly, the spacetime automorphisms are themselves infinitesimally	represented by Lie derivatives, 
                                                                                   close under another version of Lie brackets, 
																				   which then provides another notion of observables. 
I.e., once we have spacetime by 5), we can replay the Lie moves of 1) to 4) which still make sense -- 1), 3) and 4) -- in the ensuing spacetime arena to deal with 6). 
7) then conversely splits spacetime, which can be modelled with Lie derivatives, to pose 7), 
and then uses a property of 5)'s GR-constraints-specific Lie brackets algebraic structure to resolve 7).  
Moreover, the constrained observables defining brackets can be re-expressed as PDEs to which Lie's flow method can be applied, 
as a slightly extended version of a subcase of Lie's integral method for finding geometrical invariants. 

\m 

\n This picture moreover has the benefit that {\sl a lot more mathematicians and scientists} know Lie's work than 
Barbour's works on 1), 2) Dirac's on 3), 4), spacetime foliation mathematics, and so on.  
Grounding all of these works on Lie's work conveys that, once the dust has settled, our local resolution of the Problem of Time 
-- and consequent Local Theory of Background Independence -- involves familiar and elsewhere widely applicable mathematics.
This identifies a conceptual and technical starting point for understanding what the Problem of Time and Background Independence are really about. 
This starting point -- Lie's mathematics -- moreover will in many cases already be known to the reader, as well as being widely useful 
in Differential Geometry, Symmetry and Dynamics applications throughout the STEM subjects.  
In this way, our mathematical observation is moreover of pedagogical value. 

\m 

\n There is the further matter of mathematical technique versus the arena that it is used in.  
The multi-faceted Problem of Time began to be posed in detail in 1967 \cite{Battelle, DeWitt67}, in the context of QM versus GR incompatibility, 
and took until Isham and Kucha\v{r}'s reviews in the early 1990s \cite{K92, I93} to conceptually classify.  

\m 

\n On the one hand, aside from Lie's mathematics, one needs to know that QM points to 'configurations, configuration spaces, momenta, Poisson brackets, phase space, Hamiltonians, 
constraints and observables' as useful classical precursors.\footnote{QM was established in the late 1920's. 
Furthermore, it took Dirac until the 1950's to understand constraints and observables, and it took almost everyone else some decades to catch up with these developments. 
Finally, it took even longer for phase space's potential to be unlocked, particularly in a geometrical manner or as regards constraints, let alone for both of these 
to be handled concurrently in a physically satisfactory way.}
%
Remarkably, all of these bar Hamiltonians are already-TRi rather than requiring TRi modification 
(and the current Article moreover discusses how the small modification to classical Hamiltonians is of no subsequent quantum consequence).  
QM moreover itself then itself involves \cite{RS, Ringrose, I84, ABook} kinematical operators, constraint operators, observables operators. 
While the ensuing operator algebras in many ways lie beyond the remit of Lie's own mathematics, 
the quantum commutator involved in these algebras is still mathematically a Lie bracket.

\m 

\n On the other hand, one also needs to know SR (1905), GR (1915), the dynamical structure of GR: constraints, ADM formulation, 
that Hamiltonians and Poisson brackets provide a systematic way of handling constraints (from the 1950's and only consolidated in the early 1960's \cite{ADM, Dirac}).
\n GR also introduced the spacetime versus space--configuration-space--dynamics--phase-space--or--canonical dilemma \cite{ADM, Dirac, Battelle}.
By this, one can moreover consider spacetime versions of brackets and derivative concepts in place of canonical ones.
Relativity also necessitated moving away from the real-analytic functions that Lie worked with, since these do not respect causality \cite{H23, CH2}.   
This was taken into account for GR's PDEs in e.g.\ \cite{B52, HKM} using $\cC^k$, and then Sobolev, function spaces. 
GR's constraints moreover close as a Lie algebroid rather than a Lie algebra -- the Dirac algebroid -- which is named by analogy with Lie algebras 
rather than lying within the remit of Lie's own mathematics.  

\m 

\n From a pedagogical point of view, much of the previous paragraph is common knowledge among Theoretical Physicists, especially those (part-)specializing in GR.  
The most likely exceptions are, on the one hand, quantum operators, for which a background in Functional Analysis is probably more useful than one in Theoretical Physics. 
On the other hand, GR's Dirac algebroid arises, research into the mathematics of which is probably still in its infancy.

\subsection{Toward a Global Theory of Background Independence}

\n The (generalized) Killing equation's ellipticity is globally sensitive, by which its solutions the automorphism groups pick up topological information about the underlying space. 
In our physical application, this means that the infinitesimal generators that we correct our action's velocities or changes by are in fact underpinned by global features 
of the ab initio absolute space used. 

\m 

\n We moreover have a Scylla--Charybdis type dilemma: both alternatives are difficult in different ways. 

\m 

\n Scylla consists of `stratification arising from symmetry'. 
For a symmetric absolute space, the usual outcome of quotienting out the automorphism group is a stratified manifold
Its strata moreover retain some topology and symmetry information about this choice of absolute space. 
Thus such quotienting fails in its relationally desirable intent to free the modelling of all trace of absolute space.

\m 

\n Charybdis consists of `computational bereftness arising from asymmetry'. 
This follows from so many of our computational tools being rooted in symmetry 
(e.g.\ \cite{ABook} conservation, perturbation expansions, special functions, special forms taken by metrics...)

\m 

\n On the one hand, Scylla is particularly formidable when merely-Kolmogorov \cite{GT09, KKH16}, rather than Hausdorff, stratified spaces ensue. 
On the other hand, the asymmetric models leading to Charybdis are moreover generic, in the sense described in Article I, rooted on generalized Killing equations.
Thereby, nontrivial Kendall-type Relational Theories are of measure zero on the set of all possible absolute space manifolds.
This is a useful observation given recent proliferation of at least partly solvable models in this subject 
\cite{Kendall, Bhatta, FileR, AMech, PE16, I, II, III, ACirc, Minimal-N, ASphe, ATorus, Forth}.  
The trivial shape theories thereupon involve mere product spaces rather than nontrivial quotients, and thus are free of stratification or other non-manifoldness, 
for the reasons given in Article II.   
In the GR case, while the generic reduced configuration space -- Wheeler's superspace -- remains a quotient space by the corresponding automorphisms -- diffeomorphisms -- 
this remains free of strata since generic topological manifolds admit solely the trivial isotropy group. 
So the incipient Geometrodynamics of the 1960's \cite{DeWitt67, DeWitt70, Fischer70} and 1980's Kendall Shape Theory \cite{Kendall84, Kendall} went Scylla's way. 
Three decades later, Geometrodynamics \cite{FM96} contemplated Charybdis instead. 
Similarly, the incipient Shape Theory -- now of the 1980's \cite{Kendall84, Kendall89} went Scylla's way, 
whereas now, once again three decades later, proceeding via Charybdis has been contemplated as wall.

\m 

\n For arbitrary carrier space, Configurational Relationalism occurs but nongenerically. 
If it does occur, most cases are badly-behaved (merely-Kolmogorov) and are discarded as unphysical, but some cases survive.  
If it does not occur, Background Independence induces one less Problem of Time facet in the corresponding models. 
Note that this does not apply to GR version, where there is still a momentum constraint to solve even when there is no stratification.  
What does occur in this case is a globally better behaved Configurational Relationalism resolution. 
Namely, one of Bartnik and Fodor's two locality criteria for the GR Thin Sandwich -- avoiding metrics with Killing vectors, which amounts to avoiding strata -- disappears. 
There is however still no known concrete method for solving the Thin Sandwich globally in this case.  

\m 

\n Summing over all carrier or absolute spaces in Shape Theory or Relational Mechanics \cite{ABook}, or over all spatial topologies in Geometrodynamics \cite{GH92} moreover 
in principle succeeds in the relational quest of erasing all memory.
It is however both computationally expensive and computationally hard to perform (the generic entities in the sum are asymmetric).

\subsection{Localization justification of Problem of Time resolution}

\n We have argued that Hausdorff paracompact (HP) r-spaces (i.e.\ reduced or relational configuration spaces) 
are particularly amenable to justifying and combining multiple local conditions, whether in space, r-space or spaces thereover. 
For these benefit, from, firstly, partitions of unity and bump functions.
These in turn support integration, differential structure and PDEs, and thus variational principles and differential laws as we are accustomed to in Physics. 
Secondly, from possessing a Shrinking Lemma.  
Thirdly, from having a particularly straightforward theory of sheaves, by sheaf cohomology reducing to the simpler \v{C}ech cohomology for HP spaces. 
(Co)(pre)sheaves then form a solid platform for `spaces thereover', as exemplified in the current Article by constraint and observables algebraic structures.  

\m 

\n In light of Article II's considerations, let us moreover qualify staying inside a single stratum as an excision strategy, 
by which it is relationally undesirable in the physical and geometrical, if not necessarily statistical, context.

\subsection{Comparative Theory of Background Independence}

\n The current series of Articles' contribution in this direction is at the global level. 
Insofaras all Relational Mechanics considered to date have a similar local structure, and GR's conformal superspace has very similar topological properties to GR's superspace. 

\m 

\n Let us present this by paraphrasing a common theme arising in discussions between the Author and Julian Barbour. 
Upon revealing my working on multiple notions of shape and corresponding Shape Theories, his response was along the lines of 
`you are looking at the whole forest, but I like this tree here', with reference to the Similarity Shape Theory on $\mathbb{R}^d$.  
To which my original response was along the lines of `I will keep on looking at the whole forest.
Incidentally this tree you singled out in 2003 \cite{B03} is the same as the one that Kendall singled out in 1984 \cite{Kendall84}.'
[For I had already used this to fully detail the reduced configuration spaces corresponding to \cite{B03} in $\leq 2$-$d$ \cite{FORD}.]  
I now moreover add the following. 

\m 

\n `Looking at the whole forest, Geometry, Topology and a bit of Order Theory do between them highly privilege this tree of yours (and Kendall's).
In addition to this reselection, moreover, my study has shown the forest to be nongenerically small, and containing other -- if not that many -- similarly-privileged trees, 
including the following (in each case modulo isometries).   

\m 

\n 1) the $d$-tori $\mathbb{T}^d$, whose reduced configuration spaces which can readily be solved \cite{ATorus} in all dimensions, 
in contrast to the Euclidean $N$-body problem, for which $d = 1, 2$ are easy but $d \geq 3$ presents impasses.  

\m 

\n 2) The $d$-spheres $\mathbb{S}^d$ which was Kendall's \cite{Kendall} second tree since 1987 \cite{Kendall87}, for which I provided a Relational Mechanics in \cite{ASphe}. 

\m 

\n 3) The $d$-dimensional real-projective spaces $\mathbb{RP}^d$.  

\m 

\n Moreover, these other examples and, indeed, almost every other example, have no proper similarity Killing vector.
So the idea that removing scale is somehow significant is very much restricted to flat space.  

\m 

\n Considering non-maximality privilege brings in the further options of 

\m 

\n i) $Tr$  alone, with Mechanics dating back to Lagrange, 

\m 

\n i) $Dil$ alone, 

\m 

\n ii) $Rot$ alone, and 

\m 

\n iii) $Rot \times Dil$ alone \cite{AMech}, 

\m 

\n iv) $Tr \rtimes Dil$ -- Kendall's preshape sphere \cite{Kendall84}, whose corresponding Relational Mechanics was given in \cite{FileR} -- and 

\m 

\n v) $Tr \rtimes Rot$: the original Leibnizian setting of the relational side of the Absolute versus Relational (Motion) Debate, 
                  for which \cite{BB82} provided a Relational Mechanics (reviewed in \cite{FileR}).  												  

\m 
				  
\n Finally, quotienting out automorphism groups fails to remove all traces of absolute space's topology and symmetry. 
And some of the ways of dealing with this involve considering generic trees (for which this is a non-issue), 
or the sum over the whole forest, to which all trees, special or not, have a priori an opportunity to contribute to.'

\subsection{Mathematics beyond Lie toward a global resolution of the Problem of Time}

\n What our local resolution does not cover is quantum nonuniqueness issues (Multiple Choice Problems of Time) and global issues (Global Problems of Time).  
Globality requires bundles, sections, cohomology, stratified manifolds, sheaves, and a range of patching methods and of Global Analysis.  
It is here that rather more recent mathematics than Lie's work is required.   
We extend by the below names the list of mathematicians since Lie on whose work our Local Resolution of the Problem of Time is founded
It is common knowledge that Topology is due      to Poincar\'{e}, 
                            cohomology           to de Rham, 
                            fibre bundles        to Whitney, 
                            stratified manifolds to Whitney and Thom, 
                            and sheaves          to Leray, Henri Cartan and Grothendieck.  
One should also distinguish between Topology and the furtherly equipped Differential Topology -- more immediately suitable for physical applications -- 
of which cohomology is a significant part, but so are Sard's Theorem, Whitney's Theorems, and Morse Theory.  

\m 

\n While humankind's Global Analysis has not yet been sufficiently developed to deal with the Global Problems of Time, 
key works which can be used in this direction so far are those of Sobolev,  
																  Christodoulou and Klainerman.   
The Multiple Choice Problem may benefit from deformation of cohomologies, with Kontsevich's work probably being our best hope for now.
We shall only detail which techniques and citations this paragraph refers to in subsequent works.  

\m

\n For now, the current Article represents some `middle ground' in starting to globalize, which does clearly venture outside of most of the remit of Lie's work.  
This adds the names of Hausdorff     \cite{H14} -- for topological spaces themselves as well as for his separation axiom -- 
                       Alexandrov and Urysohn \cite{AU29} for the modern concept of compactness, 
                       Dieudonn\'{e} \cite{D44} for paracompactness, 
					   Lefschetz      \cite{L42} for the first -- normal spaces -- Shrinking Lemma, and 
                       \v{C}ech      \cite{C32} for his homology, 
			     which Steenrod      \cite{S36} dualized to cohomology,     
These underpinnings feed into the LCHP and LCHS stratified manifold work of Pflaum, 
                                                                            Sniatycki, and 
																			Kreck           outlined in Article II.  
																			
\m  				  
				  
\n{\bf Acknowledgments} I thank Chris Isham and Don Page for past discussions about configuration space topology, geometry and quantization, 
and they and Julian Barbour for past discussions of Background Independence.   
I also thank Don, Jeremy Butterfield, Enrique Alvarez, Reza Tavakol and Malcolm MacCallum for support with my career. 
I dedicate this piece to my first Instructor in Conceptual Thinking; without your lessons, this work would not have been possible.



\begin{thebibliography}{99}

\footnotesize


\bibitem{Newton}              I. Newton, {\it Philosophiae Naturalis Principia Mathematica} ({\it Mathematical Principles of Natural Philosophy}) (1686).  
%
                              For an English translation, see e.g.\ I.B. Cohen and A. Whitman (University of California Press, Berkeley, 1999).  
%
							  In particular, see the Scholium on Time, Place, Space and Motion therein.  
							  
\bibitem{L}                   G.W. Leibniz, {\it The Metaphysical Foundations of Mathematics} (University of Chicago Press, Chicago 1956) originally dating to 1715;  
			             
			       			  see also {\it The Leibnitz--Clark Correspondence}, ed. H.G. Alexander (Manchester 1956), originally dating to 1715 and 1716.   

\bibitem{Jacobi}              C.G.J. Jacobi, Lectures on Dynamics (1842-1843), published in (Reimer, Berlin 1866).   
							  
\bibitem{Cauchy}              A.-L. Cauchy, ``M\'{e}moire sur diverses Propri\'{e}t\'{e}s Remarquables des Substitutions R\'{e}guli$\grave{\me}$res ou Irr\'{e}guli$\grave{\me}$res, 
                              et des Syst$\grave{\me}$mes de Substitutions Conjugu\'{e}es" (Treatise on Various Remarkable Properties of (Ir)Regular Substitutions and Systems 
                              of Conjugate Substitutions) (1845). (Collected works of A.-L. Cauchy, Volume {\bf 9}).
						
\bibitem{Lie80}               S. Lie, ``Transformation groups" (1880); 
                              for an English translation, see e.g. M. Ackerman with comments by R. Hermann (1975).  							  
							  
\bibitem{M}                   E. Mach, {\it Die Mechanik in ihrer Entwickelung, Historisch-kritisch dargestellt} (J.A. Barth, Leipzig 1883);     
                              an English translation is {\it The Science of Mechanics: A Critical and Historical Account of its Development} Open Court, La  Salle, Ill. 1960).  

\bibitem{Lie}                 S. Lie and F. Engel, {\it Theory of Transformation Groups} Vols 1 to 3 (Teubner, Leipzig 1888-1893).  

\bibitem{Killing}             W. Killing, {\it Concerning the Foundations of Geometry}, J. Reine Angew Math. (Crelle) {\bf 109} 121 (1892). 
							 

\bibitem{Burnside}            W. Burnside, {\it Theory of Groups of Finite Order} 2nd ed. (Cambridge University Press, Cambridge 1911).  

\bibitem{H14}                 F. Hausdorff {\it Grundz\"{u}ge der Mengenlehre} (Principles of Set Theory) 
                              (Viet, Leipzig 1914, most recently republished as Vol 2 of F. Hausdorff's Collected Works in 2002). 

\bibitem{H23}                 J. Hadamard, {\it Lectures on Cauchy's Problem in Linear Partial Differential Equations} (Yale, New Haven 1923).

\bibitem{AU29}                P. Alexandrov and P. Urysohn, ``M\'{e}moire sur les Espaces Topologiques Compacts" (Treatise on Compact Topological Spaces), 
                              (Proceedings of Mathematical Sciences Section of Amsterdam Conference) {\bf Vol 14} (1929).

\bibitem{C32}                 E. \v{C}ech, ``Theorie G\'{e}n\'{e}rale de l'Homologie dans un Espace Quelconque" (General Theory of Homology in an Arbitrary Space), 
                              Fund. Math. {\bf 19} 149 (1932).

\bibitem{S36}                 N.E. Steenrod (Ph.D. Thesis: Princeton, 1936)  

\bibitem{L42}                 S. Lefschetz, {\it Algebraic Topology} (A.M.S., New York 1942).  

\bibitem{D44}                 J. Dieudonn\'{e}, ``Une G\'{e}n\'{e}ralisation des Espaces Compacts" (A Generalization of Compact Spaces), 
                              J. Math. Pures Appl. {\bf 23} 65 (1944).  

\bibitem{W46}                 H. Whitney, ``Complexes of Manifolds", Proc. Nat. Acad. Sci. USA {\bf 33} 10 (1946). 

\bibitem{Lanczos}             C. Lanczos, {\it The Variational Principles of Mechanics} (University of Toronto Press, Toronto 1949). 

\bibitem{DiracObs}            P.A.M. Dirac, ``Forms of Relativistic Dynamics", Rev. Mod. Phys. {\bf 21} 392 (1949).


\bibitem{DiracAlg1}           P.A.M. Dirac, ``The Hamiltonian Form of Field Dynamics", Canad. J. Math. {\bf 3} 1 (1951).  

\bibitem{B52}                 Y. Four$\grave{\me}$s-Bruhat, ``Th\'{e}or$\grave{\me}$me d'Existence pour Certains Syst$\grave{\me}$mes d'\'{E}quations 
                              aux D\'{e}riv\'{e}es Partielles Non Lin\'{e}aires" (Existence Theorems for Certain Nonlinear PDEs) Acta Mathematica \bf 88 \normalfont 141 (1952).  
									   
\bibitem{Yano55}              K. Yano, ``Theory of Lie Derivatives and its Applications (North-Holland, Amsterdam 1955).   

\bibitem{T55}                 R. Thom, ``Les Singularit\'{e}s des Applications Diff\'{e}rentiables" (Singularities in Differentiable Maps), 
							  Ann. Inst. Fourier (Grenoble) {\bf 6} 43 (1955).  

\bibitem{Clemence}            G.M. Clemence, ``Astronomical Time", Rev. Mod. Phys. {\bf 29} 2 (1957).  

\bibitem{DiracAlg2}           P.A.M. Dirac, ``The Theory of Gravitation in Hamiltonian Form", {\it Proceedings of the Royal Society of London} {\bf A 246} 333 (1958).
							  							  							  
					  							  
\bibitem{HY}                  J.G. Hocking and G.S. Young, {\it Topology} (Addison--Wesley, Reading MA 1961, reprinted by Dover, New York 1988).  

\bibitem{CH2}                 R. Courant and D. Hilbert, p. 440 of {\it Methods of Mathematical Physics} Vol 2 (Interscience, 1962). 

\bibitem{ADM}                 R. Arnowitt, S. Deser and C.W. Misner, ``The Dynamics of General Relativity", 
                              in {\it Gravitation: An Introduction to Current Research} ed. L. Witten (Wiley, New York 1962), arXiv:gr-qc/0405109.  

\bibitem{BSW}                 R.F. Baierlein, D.H. Sharp and J.A. Wheeler, ``Three-Dimensional Geometry as Carrier of Information about Time", Phys. Rev. {\bf 126} 1864 (1962).

\bibitem{G63}                 H.W. Guggenheimer, {\it Differential Geometry} (McGraw--Hill, New York 1963, reprinted by Dover, New York 1977).  					

\bibitem{Dirac}               P.A.M. Dirac, {\it Lectures on Quantum Mechanics} (Yeshiva University, New York 1964). 

\bibitem{WheelerGRT}          J.A. Wheeler, ``Geometrodynamics and the Issue of the Final State", 
                              in {\it Groups, Relativity and Topology} ed. B.S. DeWitt and C.M. DeWitt (Gordon and Breach, N.Y. 1964). 

\bibitem{A64}                 J.L. Anderson, ``Relativity Principles and the Role of Coordinates in Physics.", in {\it Gravitation and Relativity} 
                              ed. H-Y. Chiu and W.F. Hoffmann p.\  175 (Benjamin, New York 1964).   

\bibitem{Milnor-h}            J.W. Milnor, {\it Lectures on the h-Cobordism Theorem} (Princeton University Press, Princeton 1965). 

\bibitem{Serre-Lie}           J.-P. Serre, {\it Lie Algebras and Lie Groups} (Benjamin, New York 1965).  

\bibitem{W65}			      H. Whitney, ``Tangents to an Analytic Variety", Ann. Math. {\bf 81} 496 (1965). 
							  
\bibitem{A67}                 J.L. Anderson, {\it Principles of Relativity Physics} (Academic Press, New York 1967).  
														  
\bibitem{Battelle}            J.A. Wheeler, in {\it Battelle Rencontres: 1967 Lectures in Mathematics and Physics} ed. C. DeWitt and J.A. Wheeler (Benjamin, New York 1968).   
							 
\bibitem{DeWitt67}		      B.S. DeWitt, ``Quantum Theory of Gravity. I. The Canonical Theory.", Phys. Rev. {\bf 160} 1113 (1967). 
	
\bibitem{BO}                  E.P. Belasco and H.C. Ohanian, ``Initial Conditions in General Relativity: Lapse and Shift Formulation", J. Math. Phys. {\bf 10} 1503 (1969).
		
\bibitem{T69}                 R. Thom, ``Ensembles et Morphismes Stratifi\'{e}s" (Stratified Spaces and Morphisms), Bull. Amer. Math. Soc. (N.S.) {\bf 75} 240 (1969).
		
	
\bibitem{DeWitt70}            B.S. DeWitt, ``Spacetime as a Sheaf of Geodesics in Superspace", in {\it Relativity} (Proceedings of the Relativity Conference in 
                              the Midwest, held at Cincinnati, Ohio June 2-6, 1969), ed. M. Carmeli, S.I. Fickler and L. Witten (Plenum, New York 1970).

\bibitem{Fischer70}           A.E. Fischer, ``The Theory of Superspace", ibid. 

\bibitem{Magic}               C.W. Misner, ``Minisuperspace", in {\it Magic Without Magic: John Archibald Wheeler} ed. J. Klauder (Freeman, San Francisco 1972).

\bibitem{MT72}                V. Moncrief and C. Teitelboim, ``Momentum Constraints as Integrability Conditions for the Hamiltonian Constraint in General Relativity", 
                              Phys. Rev. {\bf D6} 966 (1972).    

\bibitem{York}                J.W. York Jr., ``Role of Conformal Three-Geometry in the Dynamics of Gravitation", Phys. Rev. Lett. {\bf 28} 1082 (1972); 

                              ``Conformally Invariant Orthogonal Decomposition of Symmetric Tensors on Riemannian Manifolds and the 
                              Initial-Value Problem of General Relativity", J. Math. Phys. {\bf 14} 456 (1973).
							  							  
\bibitem{MTW}                 C.W. Misner, K. Thorne and J.A Wheeler, {\it Gravitation} (Freedman, San Francisco 1973).
							  
\bibitem{T73}                 C. Teitelboim, ``How Commutators of Constraints Reflect Spacetime Structure", Ann. Phys. N.Y. {\bf 79} 542 (1973).   
							     
\bibitem{Y74}                 J.W. York Jr., ``Covariant Decompositions of Symmetric Tensors in the Theory of Gravitation", Ann. Inst. Henri Poincar\'{e} {\bf 21} 319 (1974).  

\bibitem{RS}                  M. Reed and B. Simon {\it Methods of Modern Mathematical Physics} Vols I and II. (Academic Press, New York 1975).  

\bibitem{HKM}                 T.J.R. Hughes, T. Kato and J.E. Marsden, ``Well-Posed Quasilinear Second-Order Hyperbolic Systems with Applications to Nonlinear Elastodynamics 
                              and General Relativity",  Arch. Rat. Mech. Anal. {\bf 63} 273 (1976).

\bibitem{Serre}               J.-P. Serre, {\it Linear Representations of Finite Groups} (Springer-Verlag, New York 1977).

\bibitem{Arnold}              V.I. Arnol'd, {\it Mathematical Methods of Classical Mechanics} (Springer, New York 1978).  
				  

\bibitem{BB82}                J.B. Barbour and B. Bertotti, ``Mach's Principle and the Structure of Dynamical Theories", Proc. Roy. Soc. Lond. {\bf A382} 295 (1982).  

\bibitem{John}                F. John, {\it Partial Differential Equations} (Springer, New York 1982). 

\bibitem{AMP}                 Y. Choquet-Bruhat, C. DeWitt-Morette and M. Dillard-Bleick, {\it Analysis, Manifolds and Physics} Vol. 1 (Elsevier, Amsterdam 1982).  

\bibitem{BT82}                R. Bott and L. Tu, {\it Differential Forms in Algebraic Topology} (Springer, New York 1982).
					
\bibitem{Ringrose}            R.V. Kadison and J.R. Ringrose, {\it Fundamentals of the Theory of Operator Algebras} (Academic Press, Orlando 1983).  
								
\bibitem{Kendall84}           D.G. Kendall, ``Shape Manifolds, Procrustean Metrics and Complex Projective Spaces", Bull. Lond. Math. Soc. {\bf 16} 81 (1984). 

\bibitem{I84}                 C.J. Isham, ``Topological and Global Aspects of Quantum Theory", 
                              in {\it Relativity, Groups and Topology {II}}, ed. B. DeWitt and R. Stora (North-Holland, Amsterdam 1984). 
					 
\bibitem{HallHaw}             J.J. Halliwell and S.W. Hawking, ``Origin of Structure in the Universe", Phys. Rev. {\bf D31}, 1777 (1985).

\bibitem{Fischer86}           A.E. Fischer, ``Resolving the Singularities in the Space of Riemannian Geometries", J. Math. Phys {\bf 27} 718 (1986).  

\bibitem{Kendall87}           D.G. Kendall, ``Further Developments and Applications of the Statistical Theory of Shape", 
							  Theory Probab. Appl. {\bf 31} 407 (1987).  	

\bibitem{BG88}                M Berger and B. Gostiaux, {\it Differentail Geometry: Manifolds, Curves and Surfaces} (Springer--Verlag, New York 1988).							  
							  
\bibitem{Witten88}            E. Witten, ``Topological Quantum Field Theory",              Comm. Math. Phys. {\bf 117} 353 (1988).  

\bibitem{Bell}                J.L. Bell, {\it Toposes and Local Set Theories} (Dover, New York 2008).    
							  								   							   
\bibitem{Kendall89}           D.G. Kendall, ``A Survey of the Statistical Theory of Shape", Statistical Science {\bf 4} 87 (1989). 
								   
\bibitem{DoD-Buckets}         J.B. Barbour, {\it Absolute or Relative Motion? Vol 1: The Discovery of Dynamics} (Cambridge University Press, Cambridge 1989); 

                              {\it Mach's principle: From Newton's Bucket to Quantum Gravity} ed. J.B. Barbour and H. Pfister (Birkh\"{a}user, Boston 1995).


\bibitem{H90}                 L. H\"{o}rmander, {\it The Analysis of Linear Partial Differential Operators} (Springer--Verlag, Berlin 1990).  

\bibitem{Stewart}             J.M. Stewart, {\it Advanced General Relativity} (Cambridge University Press, Cambridge 1991).

\bibitem{I91}                 C.J. Isham, ``Quantum Topology and Quantization on the Lattice of Topologies", Class. Quan. Grav {\bf 6} 1509 (1989);  
%
                              ``Quantization on the Lattice of Topologies, in {\it Florence 1989, Proceedings, Knots, Topology and Quantum Field Theories} 
							  ed. L. Lusanna (World Scientific, Singapore 1989);   

                              ``An Introduction To General Topology And Quantum Topology, unpublished, Lectures given at Banff in 1989 (available on the KEK archive); 			
							  
							  ``Canonical Groups And The Quantization Of Geometry And Topology", in {\it Conceptual Problems of Quantum Gravity} ed. 
                              A. Ashtekar and J. Stachel (Birkh\"{a}user, Boston, 1991); 
							  
                              C.J. Isham, Y.A. Kubyshin and P. Renteln, ``Quantum Metric Topology", in {\it Moscow 1990, Proceedings, Quantum Gravity} 
                              ed M.A. Markov, V.A. Berezin and V.P. Frolov (World Scientific, Singapore 1991);  

                              ``Quantum Norm Theory and the Quantization of Metric Topology", Class. Quant. Grav. {\bf 7} 1053 (1990).  	

\bibitem{HTBook}              M. Henneaux and C. Teitelboim, {\it Quantization of Gauge Systems} (Princeton University Press, Princeton 1992).   
							   
\bibitem{K92}                 K.V. Kucha\v{r}, ``Time and Interpretations of Quantum Gravity", 
                              in {\it Proceedings of the 4th Canadian Conference on General Relativity and Relativistic Astrophysics} 
                              ed. G. Kunstatter, D. Vincent and J. Williams (World Scientific, Singapore, 1992),
%
                              reprinted as Int. J. Mod. Phys. Proc. Suppl. {\bf D20} 3 (2011).   

\bibitem{I93}                 C.J. Isham, ``Canonical Quantum Gravity and the Problem of Time", in {\it Integrable Systems, Quantum Groups and Quantum Field Theories} 
                              ed. L.A. Ibort and M.A. Rodr\'{\i}guez (Kluwer, Dordrecht 1993), gr-qc/9210011.

\bibitem{GH92}                G.W. Gibbons and S.W. Hawking, ``Selection Rules for Topology Change", Commun. Math. Phys. {\bf 148} 345 (1992); 

                              G.W. Gibbons, ``Topology Change in Classical and Quantum Gravity", arXiv:1110.0611.  
							  							  
\bibitem{BF}                  R. Bartnik and G. Fodor, ``On the Restricted Validity of the Thin-Sandwich Conjecture", Phys. Rev. {\bf D48} 3596 (1993). 
						    								  						  							  
\bibitem{K93}                 K.V. Kucha\v{r}, ``Canonical Quantum Gravity", in {\it General Relativity and Gravitation 1992},  
                              ed. R.J. Gleiser, C.N. Kozamah and O.M. Moreschi (Institute of Physics Publishing, Bristol 1993), gr-qc/9304012.

\bibitem{SS94}                M. Szydlowski and J. Szczesny, ``Invariant Chaos in Mixmaster Cosmology", Phys. Rev. {\bf D50} 819 (1994). 

\bibitem{B94I}                J.B. Barbour, ``The Timelessness of Quantum Gravity. I. The Evidence from the Classical Theory", CQG {\bf 11} 2853 (1994).

\bibitem{Bala-Combi}          V.K. Balakrishnan, {\it Combinatorics} (McGraw--Hill, New York 1995).  

\bibitem{Small}               C.G.S. Small, {\it The Statistical Theory of Shape} (Springer, New York, 1996).  

\bibitem{FM96}                A.E. Fischer and V. Moncrief, 
                              ``The Reduced Hamiltonian of General Relativity and the $\sigma$-Constant of Conformal Geometry, in {\it Karlovassi 1994, 
                              Proceedings, Global Structure and Evolution in General Relativity} ed. S. Cotsakis and G.W. Gibbons 
							  (Lecture Notes in Physics, volume {\bf 460}) (Springer, Berlin 1996);     
							  
							  ``A Method of Reduction of Einstein's Equations of Evolution and a Natural Symplectic Structure on the Space of Gravitational Degrees of Freedom", 
							  Gen. Rel. Grav. {\bf 28}, 207 (1996).
							  							  
\bibitem{Kendall}             D.G. Kendall, D. Barden, T.K. Carne and H. Le, {\it Shape and Shape Theory} (Wiley, Chichester 1999).    
		
						
\bibitem{Munkres}             J.R. Munkres, {\it Topology} (Prentice--Hall, Upper Saddle River, New Jersey 2000).	

\bibitem{JM00}                K.V. Mardia and P.E. Jupp, {\it Directional Statistics} (Wiley, Chichester 2000).

\bibitem{Gotay00}             M.J. Gotay, ``Obstructions to Quantization", in {\it Mechanics: From Theory to Computation (Essays in Honor of Juan-Carlos Sim\'{o}} 
                              ed. J. Marsden and S. Wiggins, pp.\ 171-216 (Springer, New York 2000), math-ph/9809011. 

\bibitem{BFO}                 J.B. Barbour, B.Z. Foster and N. \'{o} Murchadha, ``Relativity Without Relativity", Class. Quant. Grav. {\bf 19} 3217 (2002), gr-qc/0012089.
						
\bibitem{P00}                 M.J. Pflaum, ``Smooth Structures on Stratified Spaces" Progress in Mathematics {\bf 198} 231 (2001).

\bibitem{PflaumBook}          M.J. Pflaum, {\it Analytic and Geometric Study of Stratified Spaces}, Lecture Notes in Mathematics {\bf 1768} (Springer, Berlin 2001).  

\bibitem{Ewald}               C.-O. Ewald, {\it Hochschild Homology and De Rham Cohomology of Stratifolds}, (PhD Thesis, University of Bonn 2002), 
                              http://archiv.ub.uni-heidelberg.de/volltextserver/2918/ .
		
\bibitem{B03}                 J.B. Barbour, ``Scale-Invariant Gravity: Particle Dynamics", Class. Quant. Grav. {\bf 20} 1543 (2003), gr-qc/0211021.

\bibitem{ABFO}                E. Anderson, J.B. Barbour, B.Z. Foster and N. \'{o} Murchadha, ``Scale-Invariant Gravity: Geometrodynamics", 
                              Class. Quant. Grav. {\bf 20} 157 (2003), gr-qc/0211022.
							  
\bibitem{MacCallum}           H. Stephani, D. Kramer, M.A.H. MacCallum, C.A. Hoenselaers, and E. Herlt, 
                              {\it Exact Solutions of Einstein's Field Equations} 2nd Edition (Cambridge University Press, Cambridge 2003).

\bibitem{DeWittBook}          B.S. DeWitt, {\it The Global Approach to Quantum Field Theory} Vols 1 and 2" (Oxford University Press, New York 2003). 
							  							  
\bibitem{Nagata}              {\it Encyclopedia of General Topology} ed. K.P. Hart, J.-I. Nagata and J.E. Vaughan (Elsevier, Amsterdam 2003).  
							  
\bibitem{FH}                  W. Fulton and J. Harris, ``Representation Theory. A First Course" (Springer, New York 2004).  
							 
\bibitem{R04}                 C. Rovelli, {\it Quantum Gravity} (Cambridge University Press, Cambridge 2004).  

\bibitem{MP05}                K.V. Mardia and V. Patrangenaru, ``Directions and Projective Shapes", Annals of Statistics {\bf 33} 1666 (2005), math/0508280. 

\bibitem{Giu06}	    		  D. Giulini, ``Some Remarks on the Notions of General Covariance and Background Independence", 
                              in {\it An Assessment of Current Paradigms in the Physics of Fundamental Interactions} 
							  ed. I.O. Stamatescu, Lect. Notes Phys. {\bf 721} 105 (2007), arXiv:gr-qc/0603087.

\bibitem{FORD}                E. Anderson, ``Foundations of Relational Particle Dynamics", Class. Quant. Grav. {\bf 25} 025003 (2008), arXiv:0706.3934.   

\bibitem{FEPI}                E. Anderson, ``New Interpretation of Variational Principles for Gauge Theories. I. Cyclic Coordinate Alternative to ADM Split", 
                              Class. Quant. Grav. {\bf 25} 175011 (2008), arXiv:0711.0288.

\bibitem{Wells}               R.O. Wells, {\it Differential Analysis on Complex Manifolds} (Springer, New York 2008).  
							  							  
\bibitem{NSW08}               P. Niyogi, S. Smale and S. Weinberger, ``Finding the Homology of Submanifolds with High Confidence from Random Samples", 
                              Discrete Comput. Geom. {\bf 39} 419 (2008). 
							 													
\bibitem{Giu09}               D. Giulini, ``The Superspace of Geometrodynamics", Gen. Rel. Grav. {\bf 41} 785 (2009) 785, arXiv:0902.3923.  

\bibitem{PPSCT}               E. Anderson, ``Relational Motivation for Conformal Operator Ordering in Quantum Cosmology", Class. Quant. Grav. {\bf 27} 045002 (2010), arXiv:0905.3357.  
		
\bibitem{GT09}			      D. Groisser, and H.D. Tagare, ``On the Topology and Geometry of Spaces of Affine Shapes", 
                              Journal of Mathematical Imaging and Vision {\bf 34} 222 (2009).  
				
		
\bibitem{APoT}                E. Anderson, ``The Problem of Time in Quantum Gravity", in {\it Classical and Quantum Gravity: Theory, Analysis and Applications}  
                              ed. V.R. Frignanni (Nova, New York 2012), arXiv:1009.2157.   
		  							  	 							 							 							  
\bibitem{Bojo1}               M. Bojowald, P.A. Hoehn and A. Tsobanjan, ``An Effective Approach to the Problem of Time", Class. Quant. Grav. {\bf 28} 035006 (2011), arXiv:1009.5953.  

\bibitem{Bojo2}               M. Bojowald, P.A. Hoehn and A. Tsobanjan, ``Effective Approach to the Problem of Time: General Features and Examples", 
                              Phys. Rev. {\bf D83} 125023 (2011), arXiv:1011.3040.    

\bibitem{Kreck}               M. Kreck, {\it Differential Algebraic Topology: From Stratifolds to Exotic Spheres} (American Mathematical Society, Providence 2010).							 

\bibitem{Bob-1}               R.J. Adler, O. Bobrowski, M.S. Borman, E. Subag, and S. Weinberger, "Persistent Homology for Random Fields and Complexes",  
                              Collections {\bf 6} 124 (2010).  
 							  
\bibitem{FileR}               E. Anderson, ``The Problem of Time and Quantum Cosmology in the Relational Particle Mechanics Arena", arXiv:1111.1472. 

\bibitem{Bojo3}               P.A. Hoehn, E. Kubalova and A. Tsobanjan, ``Effective Relational Dynamics of a Nonintegrable Cosmological Model", 
                              Phys. Rev. {\bf D86} 065014 (2012), arXiv:1111.5193.
							  
\bibitem{Bojowald}            M. Bojowald, {\it Canonical Gravity and Applications: Cosmology, Black Holes, and Quantum Gravity} (CUP, Cambridge 2011).   

\bibitem{APoT2}				  E. Anderson, ``Problem of Time in Quantum Gravity", Annalen der Physik, {\bf 524} 757 (2012),  arXiv:1206.2403.    

\bibitem{Tene}                H. Tene, {\it Some Geometric Equivariant Cohomology Theories} (PhD Thesis, University of Bonn 2010), arXiv:1210.7923.  
							  			
\bibitem{Bhatta}              A. Bhattacharya and R. Bhattacharya, {\it Nonparametric Statistics on Manifolds with Applications to Shape Spaces} 
                              (Cambridge University Press, Cambridge 2012).
							 							 				
\bibitem{Lee2}                J.M. Lee, {\it Introduction to Smooth Manifolds} 2nd Ed. (Springer, New York 2013).
									
\bibitem{SniBook}             J. \'{S}niatycki, {\it Differential Geometry of Singular Spaces and Reduction of Symmetry} (Cambridge University Press, Cambridge 2013).
							 							 
\bibitem{AM13}                E. Anderson and F. Mercati, ``Classical Machian Resolution of the Spacetime Construction Problem", arXiv:1311.6541. 
 					 							 
\bibitem{ABeables}            E. Anderson,  ``Beables/Observables in Classical and Quantum Gravity", SIGMA {\bf 10} 092 (2014), arXiv:1312.6073.  

\bibitem{APoT3}               E. Anderson, ``Problem of Time and Background Independence: the Individual Facets", arXiv:1409.4117, 
                              a more advanced version of which comprises Chapters 8, 9, 10 and 12 of \cite{ABook}.    

\bibitem{Ghrist}              R. Ghrist, {\it Elementary Applied Topology} (2014). 

\bibitem{ASoS}                E. Anderson, ``Spaces of Spaces", arXiv.1412.0239.

\bibitem{TRiPoD}              E. Anderson, ``TRiPoD (Temporal Relationalism implementing Principles of Dynamics)", arXiv:1501.07822.  

\bibitem{AMech}               E. Anderson, ``Six New Mechanics corresponding to further Shape Theories", Int. J. Mod. Phys. {\bf D 25} 1650044 (2016), arXiv:1505.00488. 

\bibitem{ASphe}               E. Anderson, ``Spherical Relationalism", arXiv:1505.02448; 

                                           ``Relational Theory on $\mathbb{S}^2$", forthcoming February 2019.   
 
\bibitem{AObs2}               E. Anderson,  ``Explicit Partial and Functional Differential Equations for Beables or Observables" arXiv:1505.03551.  

\bibitem{TRiFol}              E. Anderson,  ``Problem of Time: Temporal Relationalism Compatibility of Other Local Classical Facets Completed", arXiv:1506.03517.

\bibitem{KT15}                M. Kreck and H. Tene, ``Hilbert Stratifolds and a Quillen Type Geometric Description of Cohomology for Hilbert Manifolds", arXiv:1506.07075.   

\bibitem{AObs3}               E. Anderson,  ``On Types of Observables in Constrained Theories", arXiv:1604.05415.   

\bibitem{DM16}                I.L. Dryden, K.V. Mardia, {\it Statistical Shape Analysis}, 2nd Edition (Wiley, Chichester 2016).  
	
\bibitem{PE16}                V. Patrangenaru and L. Ellingson, ``Nonparametric Statistics on Manifolds and their Applications to Object Data Analysis" 
                             (Taylor and Francis, Boca Raton, Florida 2016).  

\bibitem{KKH16}               F. Kelma, J.T. Kent and T. Hotz, ``On the Topology of Projective Shape Spaces", arXiv:1602.04330. 

\bibitem{Wedhorn}             T. Wedhorn, {\it Manifolds, Sheaves, and Cohomology} (Springer, Wiesbaden 2016).
							 
\bibitem{ABook}               E. Anderson, {\it The Problem of Time. Quantum Mechanics versus General Relativity} , Fundam. Theor. Phys. {\bf 190} (Springer, 2017).  

\bibitem{Epi-B}               See Epilogues II.B and III.B of \cite{ABook}.  
	            
\bibitem{Epi-C}               See Epilogues II.C and III.C of \cite{ABook}. 
		
\bibitem{Bob-2}               O. Bobrowski and S. Weinberger, "On the Vanishing of Homology in Random \v{C}ech Complexes", Random Structures and Algorithms {\bf 51} 14--51 (2017).
			
\bibitem{I}                   E. Anderson, ``The Smallest Shape Spaces. I. Shape Theory Posed, with Example of 3 Points on the Line", arXiv:1711.10054.  
						  							  
\bibitem{II}                  E. Anderson, ``The Smallest Shape Spaces. II. 4 Points on a Line Suffices for a Complex Background-Independent Theory of Inhomogeneity", 
										     arXiv:1711.10073. 

\bibitem{III}                 E. Anderson, ``The Smallest Shape Spaces. III. Triangles in the Plane and in 3-$d$", arXiv:1711.10115. 

\bibitem{MBook}               F. Mercati {\it Shape Dynamics: Relativity and Relationalism} (O.U.P., New York 2018).

\bibitem{A-Monopoles}         E. Anderson, ``Monopoles of Twelve Types in 3-Body Problems", arXiv:1802.03465.

\bibitem{ACirc}				  E. Anderson, ``Background Independence: $\mathbb{S}^1$ and $\mathbb{R}$ Absolute Spaces Differ Greatly in Shape-and-Scale Theory", arXiv:1804.10933. 
							  
\bibitem{Top-Shapes}          E. Anderson, ``Topological Shape Theory", arXiv:1803.11126; 

                                           ``Rubber Relationalism: Smallest Graph-Theoretically Nontrivial Leibniz Spaces", arXiv:1805.03346.  

\bibitem{A-Generic}           E. Anderson  ``Absolute versus Relational Debate: a Modern Global Version", arXiv:1805.09459.
										   
\bibitem{Minimal-N}           E. Anderson, ``$N$-Body Problem: Minimal $N$ for Qualitative Nontrivialities", arXiv:1807.08391; 

  							               ``$N$-Body Problem: Minimal $N$ for Qualitative Nontrivialities II: Varying Carrier Space and Quotiented Group", forthcoming 2019.  
					
\bibitem{A-Lett}	          E. Anderson, ``A Local Resolution of the Problem of Time", arXiv:1809.01908.     
					
\bibitem{PE-1}                E. Anderson, ``Specific PDEs for Preserved Quantities in Geometry. I. Similarities and Subgroups", arXiv:1809.02045.  

\bibitem{DO-1}                E. Anderson, ``Spaces of Observables from Solving PDEs. I. Translation-Invariant Theory.", arXiv:1809.07738. 

\bibitem{IV-2}                E. Anderson, ``Quadrilaterals in Shape Theory. II. Alternative Derivations of Shape Space: Successes and Limitations", arXiv:1810.05282. 
		
\bibitem{A-Brackets}          E. Anderson,  ``Geometry from Brackets Consistency", arXiv:1811.00564.

\bibitem{A-Killing}           E. Anderson, ``Shape Theories. I. Their Diversity is Killing-Based and thus Nongeneric", arXiv:1811.06516.  

\bibitem{A-Cpct}              E. Anderson, ``Shape Theories II. Compactness Selection Principles", arXiv:1811.06528. 

\bibitem{Event-Shapes}		  E. Anderson,  ``Event Shapes", forthcoming January 2019.
				
\bibitem{ATorus}              E. Anderson, ``Background Independence and Shape-and-Scale Theory on the Torus, forthcoming January 2019.   

\bibitem{Forth}               E. Anderson, forthcoming.   
	            			   	
\end{thebibliography}
\end{document}